\documentclass[12pt,preprint]{aastex}
\usepackage[onecolumn,numberedappendix]{emulateapj5}
\usepackage{epsfig}
\usepackage{multirow}
\usepackage{graphics}
\usepackage{amsmath, amsthm, amssymb}
\tightenlines
\newcommand \lsim{\mathrel{\rlap{\lower4pt\hbox{\hskip1pt$\sim$}}
    \raise1pt\hbox{$<$}}}
\newcommand \gsim{\mathrel{\rlap{\lower4pt\hbox{\hskip1pt$\sim$}}
    \raise1pt\hbox{$>$}}}

\newcommand     \kms    {\,{\rm km~s}^{-1}}

\newcommand{\beq}{\begin{equation}}
\newcommand{\eeq}{\end{equation}}
\newcommand{\beqa}{\begin{eqnarray}}
\newcommand{\eeqa}{\end{eqnarray}}

\newcommand{\thco}	{^{13}{\rm CO}}

\newcommand{\Sigco}     {\Sigma_{13\rm{CO}}}
\newcommand{\Sigsmf}     {\Sigma_{\rm SMF}}

\newlength{\figwidth}
\begin{document}

\title{The Dynamical State of Filamentary Infrared Dark Clouds}

%\centerline{DRAFT: \today}
\author{Audra K. Hernandez}
\affil{Department of Astronomy, University of Florida, Gainesville, FL 32611, USA;\\ audrah@astro.ufl.edu}
\author{Jonathan C. Tan}
%\affil{Department of Astronomy, University of Florida, Gainesville, FL 32611, USA; audrah@astro.ufl.edu, jt@astro.ufl.edu}
%emulate
\affil{Departments of Astronomy \& Physics, University of Florida, Gainesville, FL 32611, USA;\\ jt@astro.ufl.edu}

\begin{abstract}
The dense, cold gas of Infrared Dark Clouds (IRDCs) is thought to be
representative of the initial conditions of massive star and star
cluster formation. We analyze $\thco$ $J=1-0$ line emission data from
the Galactic Ring Survey of Jackson et al. for two filamentary IRDCs,
comparing the mass surface densities derived from $\thco$, $\Sigco$,
with those derived from mid-infrared small median filter extinction
mapping, $\Sigsmf$, by Butler \& Tan. After accounting for molecular
envelopes around the filaments, we find approximately linear relations
between $\Sigco$ and $\Sigsmf$, i.e. an approximately constant ratio
$\Sigco/\Sigsmf$ in the clouds. There is a variation of about a factor
of two between the two clouds. We find evidence for a modest decrease
of $\Sigco/\Sigsmf$ with increasing $\Sigma$, which may be due to a
systematic decrease in temperature, increase in importance of high
$\thco$ opacity cores, increase in dust opacity, or decrease in
$\thco$ abundance due to depletion in regions of higher column
density. We perform ellipsoidal and filamentary virial analyses of the
clouds, finding that the surface pressure terms are dynamically
important and that globally the filaments may not yet have reached
virial equilibrium. Some local regions along the filaments appear to
be close to virial equilibrium, although still with dynamically
important surface pressures, and these appear to be sites where star
formation is most active.
%, i.e. no evidence for depletion up to
%$\Sigma \simeq 0.05\:\gcc$, equivalent to $N_{\rm H} \simeq 2.1\times
%10^{22}\:{\rm cm^{-2}}$ or $A_V\simeq 10$~mag. However, the ratio of
%$\Sigco$ and $\Sigsmf$ varies between the two filaments by a factor of
%3, which could be due to abundance variations at this level and/or
%temperature differences of $\sim 10$~K. We consider the kinematics of
%the IRDC filaments, finding they are in approximate virial balance.
\end{abstract}


\section{Introduction}

Massive, high column density Infrared Dark Clouds (IRDCs), typically
identified as being opaque against the Galactic background at $\sim 10
{\rm \rm{\mu} m}$, are thought to contain the sites of future massive
star and star cluster formation (e.g. Rathborne et al. 2006), since
their densities ($n_{\rm H}\gsim 10^4\:{\rm cm^{-3}}$) and mass
surface densities ($\Sigma \gsim 0.1\:{\rm g\:cm^{-2}}$) are similar
to regions known to be undergoing such formation activity (Tan 2007).
Studies of molecular line emission from IRDCs can help determine their
kinematics. In particular, we would like to know if they are
gravitationally bound, if they are near virial equilibrium and if
there is evidence for coherent gas motions that might indicate that
IRDC formation involves converging atomic flows (Heitsch et al. 2008)
or converging molecular flows from cloud collisions (Tan 2000).

In this study we use $\thco$ $J=1-0$ line emission data from the
Galactic Ring Survey (GRS) (Jackson et al. 2006) for two filamentary
IRDCs, clouds F ($l=34.437^\circ$, $b=0.245^\circ$, $d=3.7$~kpc) and H
($l=35.395^\circ$, $b=-0.336^\circ$, $d=2.9$~kpc) from the sample of
10 relatively nearby massive and dense IRDCs of Butler \& Tan (2009,
hereafter BT09), comparing $\thco$-derived mass surface densities,
$\Sigco$, with small median filter (SMF) mid-infrared (MIR) ($\rm 8\mu
m$) extinction mapping derived mass surface densities, $\Sigsmf$,
using the method of BT09 applied to the {\it Spitzer} Infrared Array
Camera (IRAC) band 4 images of the Galactic plane taken as part of the
Galactic Legacy Mid-Plane Survey Extraordinaire (GLIMPSE) (Benjamin et
al. 2003). We consider systematic errors in each of these methods,
which is necessary before analyzing larger samples of clouds. We are
also able to look for evidence of changing CO abundance with column
density, e.g. due to possible depletion of CO at high densities. We
then perform a virial analysis of the clouds to determine their
dynamical state.

There have been a number of other studies comparing $\thco$ derived
mass surface densities with those from other methods. For example,
Goodman, Pineda \& Schnee (2009) compared near infrared (NIR) dust
extinction, far infrared (FIR) dust emission and $\thco$ line emission
in the Perseus giant molecular cloud (GMC), probing values of $\Sigma$
up to $\sim 0.02\:{\rm g\:cm^{-2}}$ (i.e. up to $A_V\simeq
8$~mag). Even after accounting for temperature and optical depth
variations they concluded that $\thco$ emission was a relatively
unreliable tracer of mass surface density, perhaps due to threshold,
depletion and opacity effects. Our study probes higher values of
$\Sigma$, from $\sim 0.01$ to $\sim 0.05\:{\rm g\:cm^{-2}}$, and
compares $\thco$ emission with MIR extinction in order to investigate
these processes.

Battersby et al. (2010) used $\thco$ emission, MIR extinction and FIR
dust emission methods to measure $\Sigma$ and mass of clumps in 8
IRDCs, one of which is IRDC F of our study. They did not present a
specific comparison of $\Sigco$ with other methods, although derived
clump masses were in reasonable agreement. Their sample also included
MIR-bright regions, associated with ultra-compact HII regions, for
which the MIR extinction method cannot be applied. As we describe
below, our approach differs in a number of ways, including by focusing
on filamentary and mostly quiescent regions of IRDCs for which the MIR
extinction method is most reliable and which are likely to be closer
to the initial conditions of the massive star and star cluster
formation process. We note that while IRDC F in particular does
contain some regions of quite active star formation, including an
ultra-compact \ion{H}{2} region, here we have concentrated on its more
quiescent portions.

\section{Mass Surface Density Estimation From $\thco$}

\begin{figure}[!tb]
%\begin{figure*}
%\includegraphics[width=3.5in,angle=0]{TauvsTmb.eps}f1a.eps
%\includegraphics[width=3.5in,angle=0]{dN13vsTmb.eps}f1b.eps & 
\begin{center}$
\begin{array}{cc}
\includegraphics[width=3.5in,angle=0]{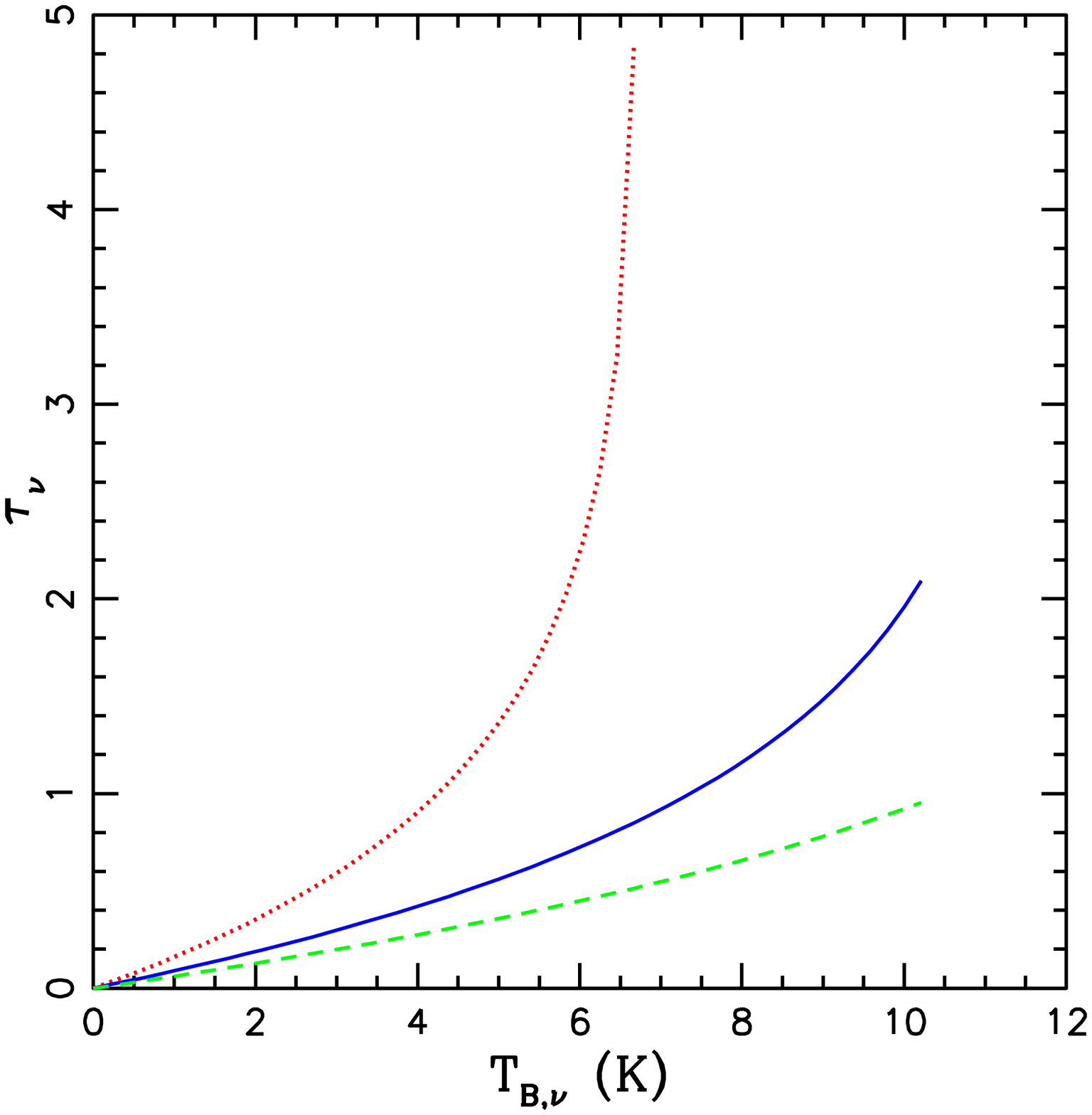}
\includegraphics[width=3.5in,angle=0]{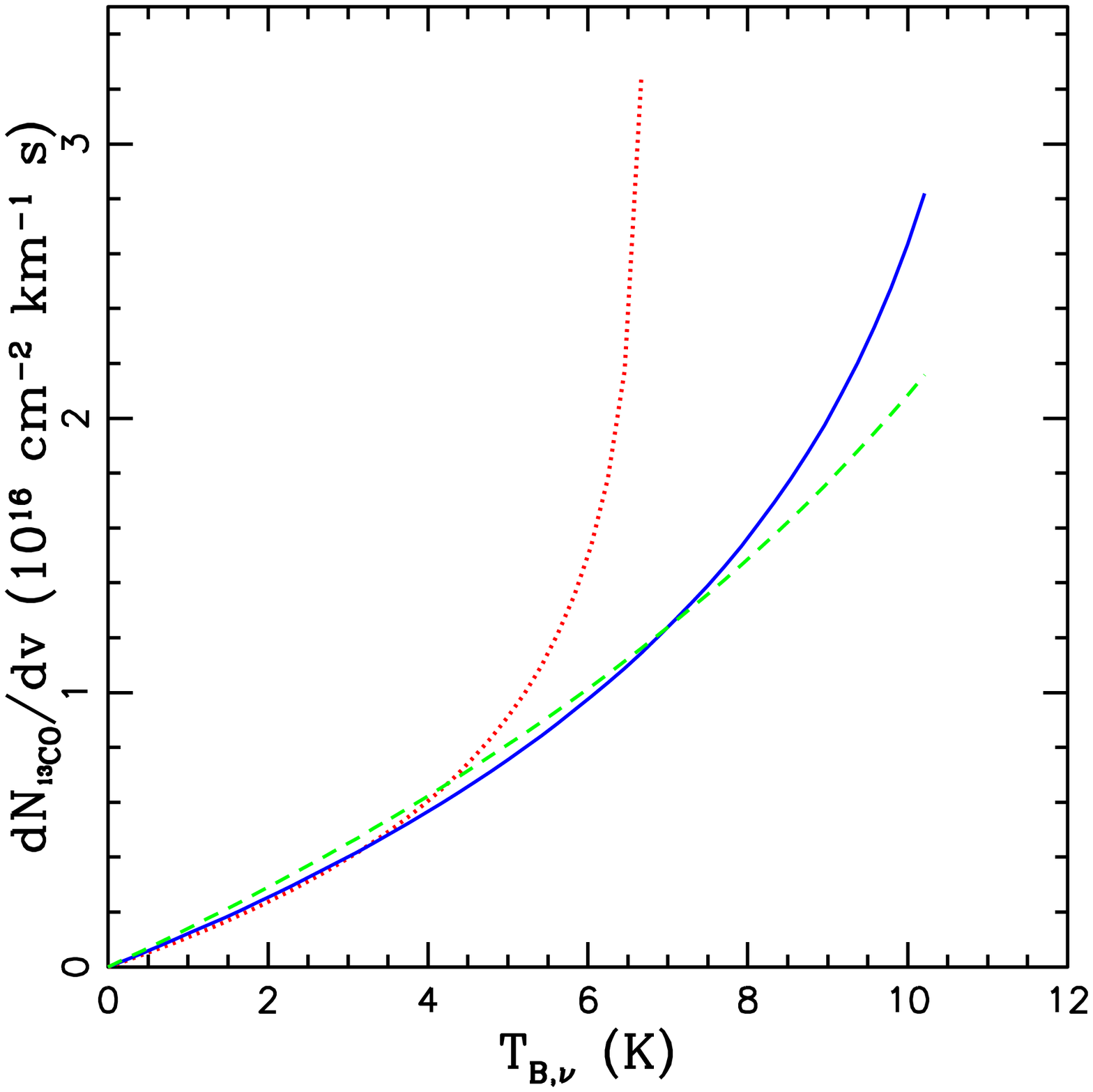} & 
\end{array}$
\end{center}
\caption{(a) Left panel: Dependence of $\tau_\nu$ with $T_{B,\nu}$ (eq. \ref{eq:detection}) assuming $T_{\rm ex}=10,15,20$~K (red dotted, blue solid, green dashed lines, respectively). (b) Right panel: Dependence of ${\rm d}N_{\rm 13CO}/{\rm d}v$ with $T_{B,\nu}$ assuming $T_{\rm ex}=10,15,20$~K (red dotted, blue solid, green dashed lines, respectively).
}
\label{fig:tauvsT}
\end{figure}

We evaluate the column density of $\thco$ molecules, ${\rm d} N_{\rm
  13CO}$, in a velocity interval ${\rm d}v$ from their $J=1\rightarrow 0$
emission via 
\beq \frac{{\rm d}N_{\rm 13CO}(v)}{{\rm d}v} = \frac{8\pi
  Q_{\rm rot}}{A \lambda_0^3} \frac{g_l}{g_u} \tau_\nu \left[1-{\rm
    exp}\left(-\frac{h\nu}{kT_{\rm ex}}\right)\right]^{-1},
\label{eq:dN}
\eeq where $Q_{\rm rot}$ is the partition function, $A = 6.3355\times
10^{-8}{\rm s^{-1}}$ is the Einstein coefficient, $\lambda_0=
0.27204$~cm, $g_l = 1$ and $g_u = 3$ are the statistical weights of
the lower and upper levels, $\tau_\nu$ is the optical depth of the
line at frequency $\nu$, i.e. at velocity $v$, $T_{\rm ex}$ is the
excitation temperature (assumed to be the same for all rotational
levels). For linear molecules, the partition function is $Q_{\rm rot}=
\sum_{J=0}^{\infty} (2J+1) {\rm exp}(-E_J/kT_{\rm ex})$ with $E_J =
J(J+1)h B$ where $J$ is the rotational quantum number and
$B=5.5101\times 10^{10}\:{\rm s}^{-1}$ is the rotational
constant. Thus for $\thco$(1-0) we have $E_J/k = 5.289$~K. For $J=1$,
$Q_{\rm rot}= 4.134,6.018,7.906$ for $T_{\rm ex}=10,15,20$~K.

The optical depth is determined via
\beq
T_{B,\nu} = \frac{h\nu}{k} [f(T_{\rm ex}) - f(T_{\rm bg})] \left[1 - e^{-\tau_\nu}\right],
\label{eq:detection}
\eeq where $T_{B,\nu}$ is the brightness temperature at frequency
$\nu$, $f(T)\equiv [{\rm exp}(h\nu/[kT])-1]^{-1}$, and $T_{\rm
  bg}=2.725$~K is the background temperature. $T_{B,\nu}$ is derived
from the antenna temperature, $T_A$, via $T_A\equiv \eta f_{\rm
  clump}T_{B,\nu}$, where $\eta$ is the main beam efficiency
($\eta=0.48$ for the GRS) and $f_{\rm clump}$ is the beam dilution
factor of the $\thco$ emitting gas, which we assume to be 1 for the
IRDCs we are studying. However, it should be noted that the BT09 MIR
extinction maps of the IRDCs do show that there is structure on scales
smaller than the angular resolution of the GRS survey. Given the
observed $T_{B,\nu}$ and for an assumed $T_{\rm ex}$ we solve
equations (\ref{eq:dN}) and (\ref{eq:detection}) for $\tau_\nu$ and
thus ${\rm d}N_{\rm 13CO}/{\rm d}v$ (see Fig.~\ref{fig:tauvsT}). For a
given $T_{\rm B,\nu}$, this figure allows us to judge the sensitivity
of the derived column density per unit velocity to temperature
uncertainties.

While we use the above formulae, which account for optical depth, to
calculate our column density estimates, for convenience we also state
their behavior in the limit of optically thin conditions. We have
$T_{B,\nu}=(h\nu/k)[f(T_{\rm ex}) - f(T_{\rm bg})]\tau_\nu$ so that
\begin{eqnarray}
\frac{{\rm d}N_{\rm 13CO}(v)}{{\rm d}v} & = & 1.242\times 10^{14} \frac{Q_{\rm rot}}{f(T_{\rm ex}) - f(T_{\rm bg})} [1-{\rm exp}(-h\nu/ k T_{\rm ex})]^{-1}\frac{T_A/K}{\eta f_{\rm clump}} \:{\rm cm^{-2} km^{-1}s}\\
 & \rightarrow & 1.144 \times 10^{15}\frac{T_A/K}{\eta f_{\rm clump}} \:{\rm cm^{-2} km^{-1}s}\:\:(T_{\rm ex}=15~{\rm K}).
\label{eq:thin}
\end{eqnarray}
The last coefficient changes to $(0.9865,1.347)\times 10^{15}$ for $T_{\rm ex}=10,20$~K.

Devine (2009) estimates a temperature of 19~K for cloud F based on VLA
observations of $\rm NH_3(1,1)$ and $\rm (2,2)$. For this cloud, we
thus adopt a temperature of 20~K. For cloud H we choose a more typical
IRDC temperature of 15~K (Carey et al. 1998; Carey et al. 2000; Pillai
et al. 2006). We also consider the effect of varying these
temperatures.

\begin{figure}[!tb]
%\begin{figure*}
%\includegraphics[width=3.5in,angle=0]{taucompF_3_20K.eps}f2a.eps & 
%\includegraphics[width=3.5in,angle=0]{taucompH_3.eps}f2b.eps
\begin{center}$
\begin{array}{cc}
\includegraphics[width=3.5in,angle=0]{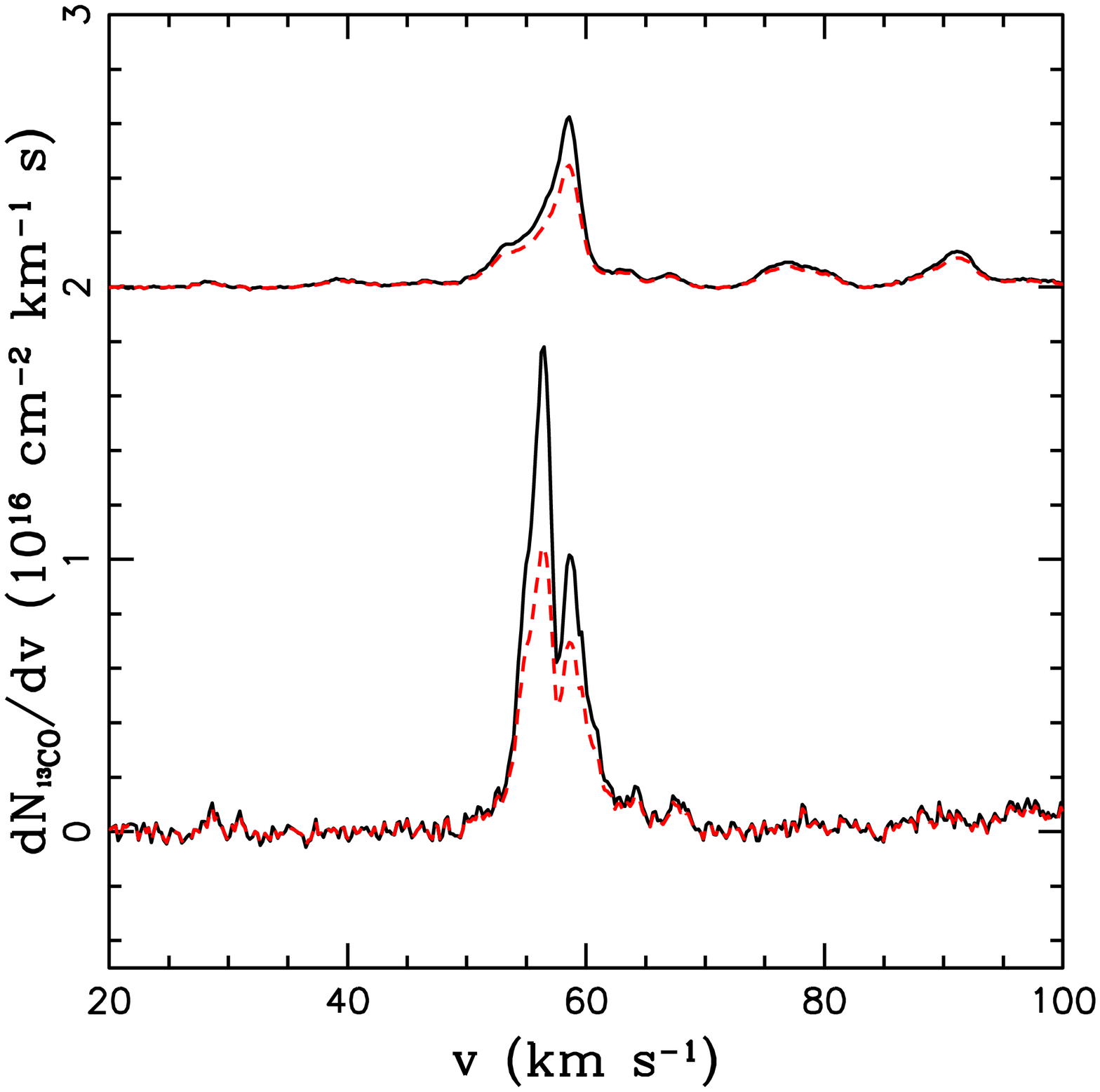} & 
\includegraphics[width=3.5in,angle=0]{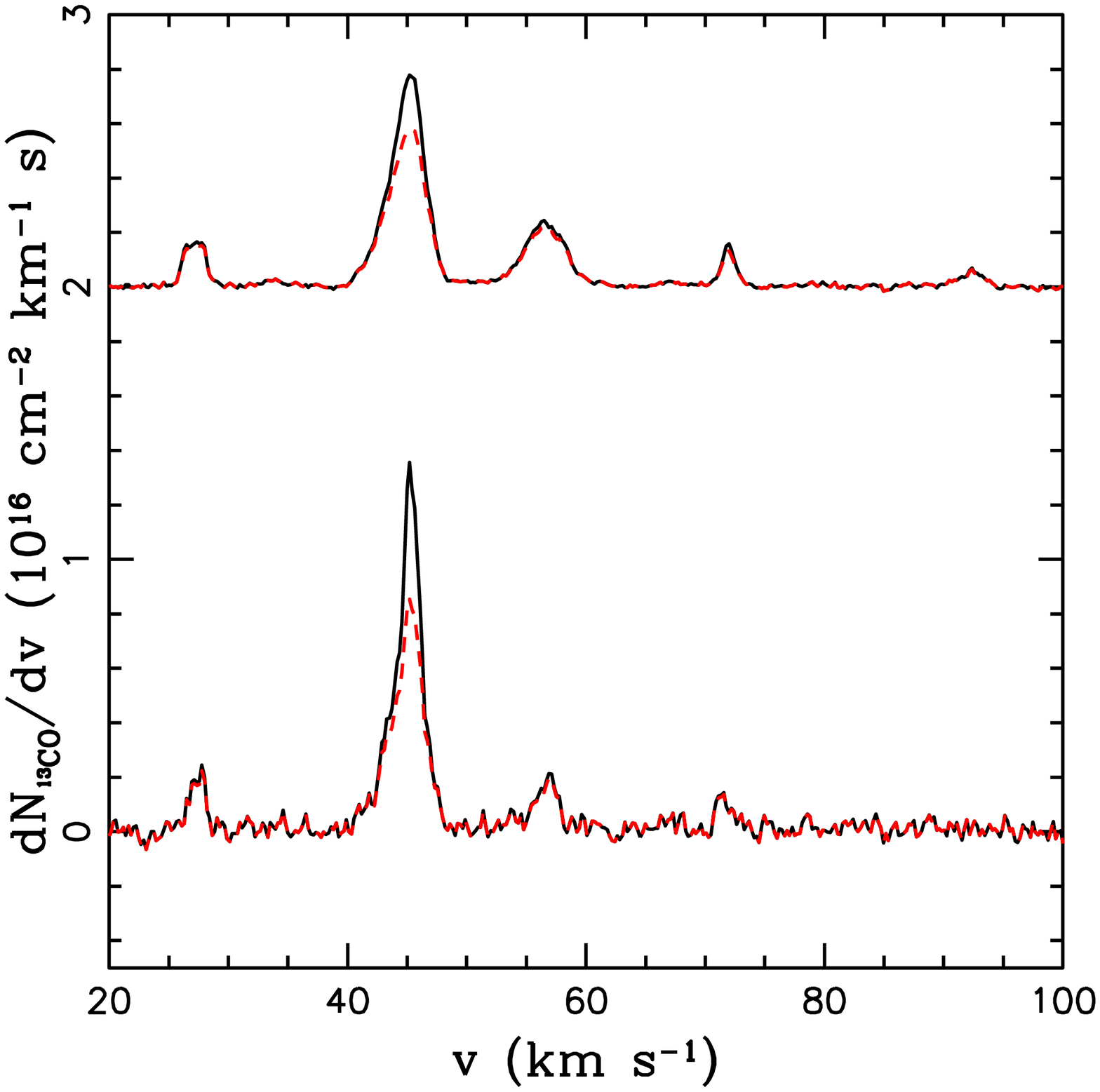}
\end{array}$
\end{center}
\caption{Column density profiles with velocity for IRDCs F (left) and
  H (right). The dashed lines shows the profiles assuming optically
  thin emission, while the solid lines show the profiles after
  correction for optical depth. A temperature of 20~K was assumed for
  IRDC F and 15~K for IRDC H (see text). For each cloud the upper,
  offset profile is that of the average for the cloud and the lower
  profile is that of the highest column density position.}
\label{spectra}
\end{figure}

\begin{figure}[!tb]
%\begin{figure*}

%\includegraphics[width=2.4in, angle=0]{f3TL.ps}f3a.ps 
%\includegraphics[width=2.4in, angle=0]{f3ML.ps}f3b.ps 
%\includegraphics[width=2.4in, angle=0]{f3BL.ps}f3c.ps 
%\\
%\includegraphics[width=2.4in, angle=0]{f3TR.ps}f3d.ps      
%\includegraphics[width=2.4in, angle=0]{f3MR.ps}f3e.ps
%\includegraphics[width=2.4in, angle=0]{f3BR.ps}f3f.ps

\begin{center}$
\begin{array}{cc}
\includegraphics[width=2.4in, angle=0]{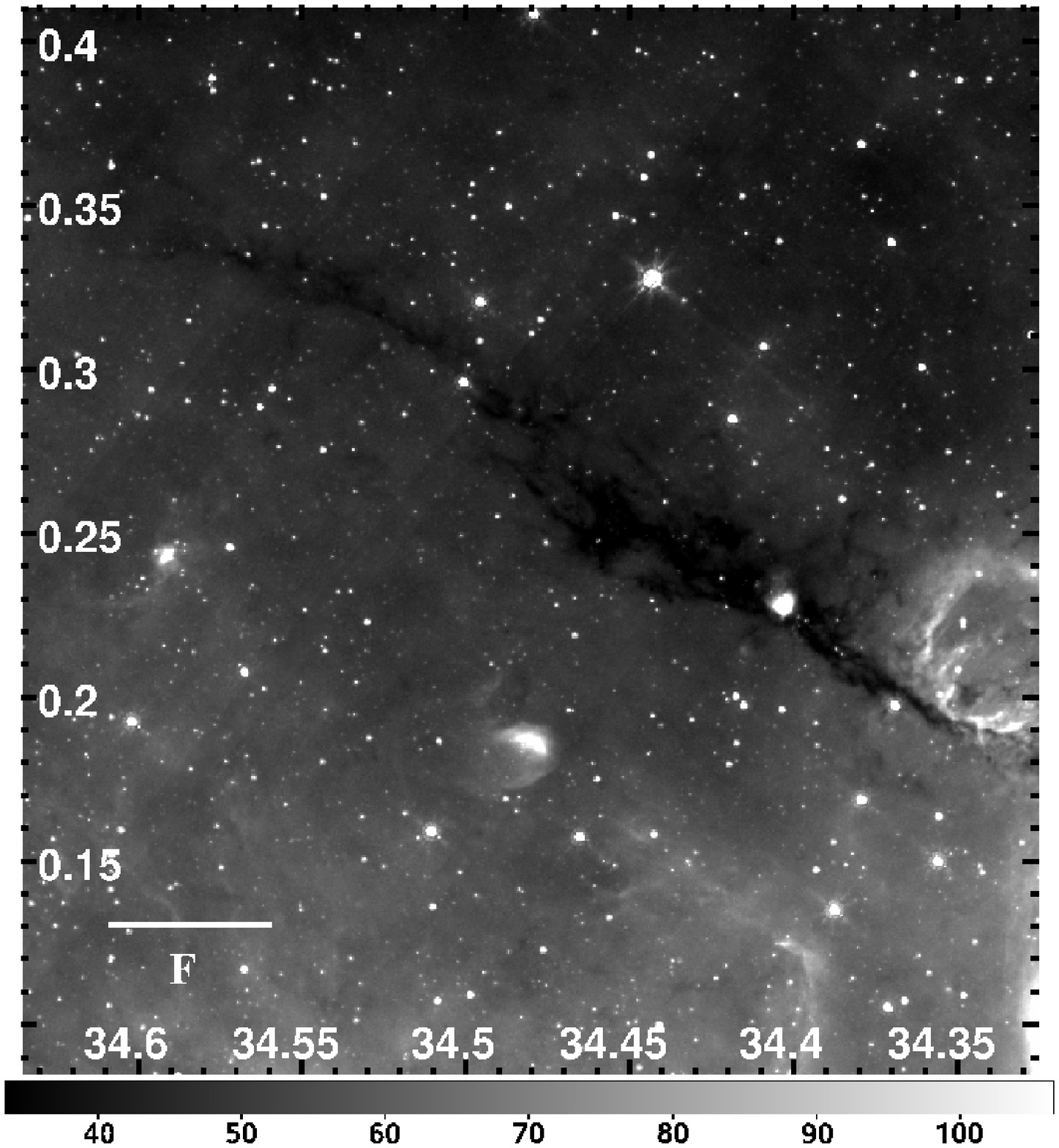} 
\includegraphics[width=2.4in, angle=0]{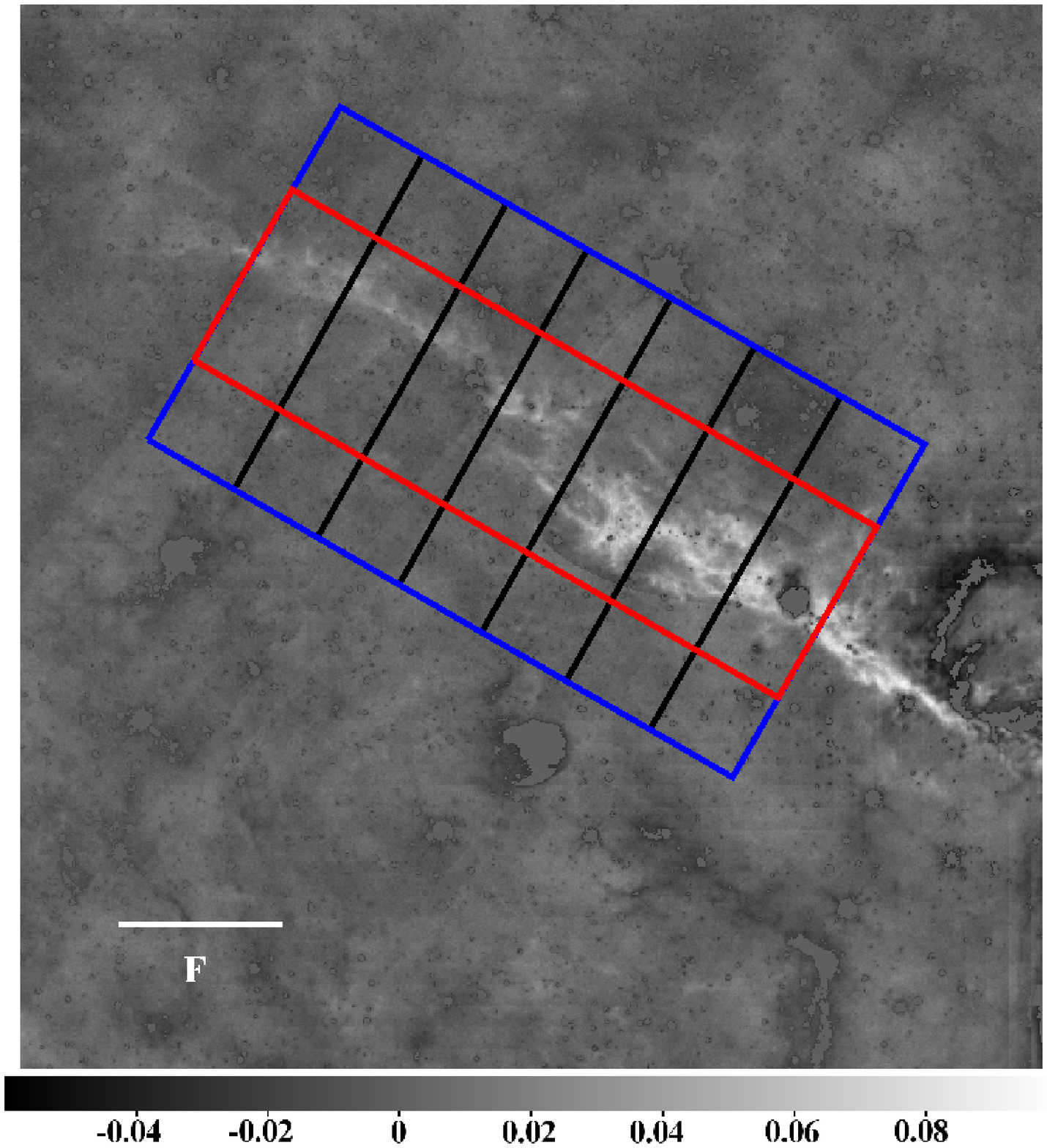} 
\includegraphics[width=2.4in, angle=0]{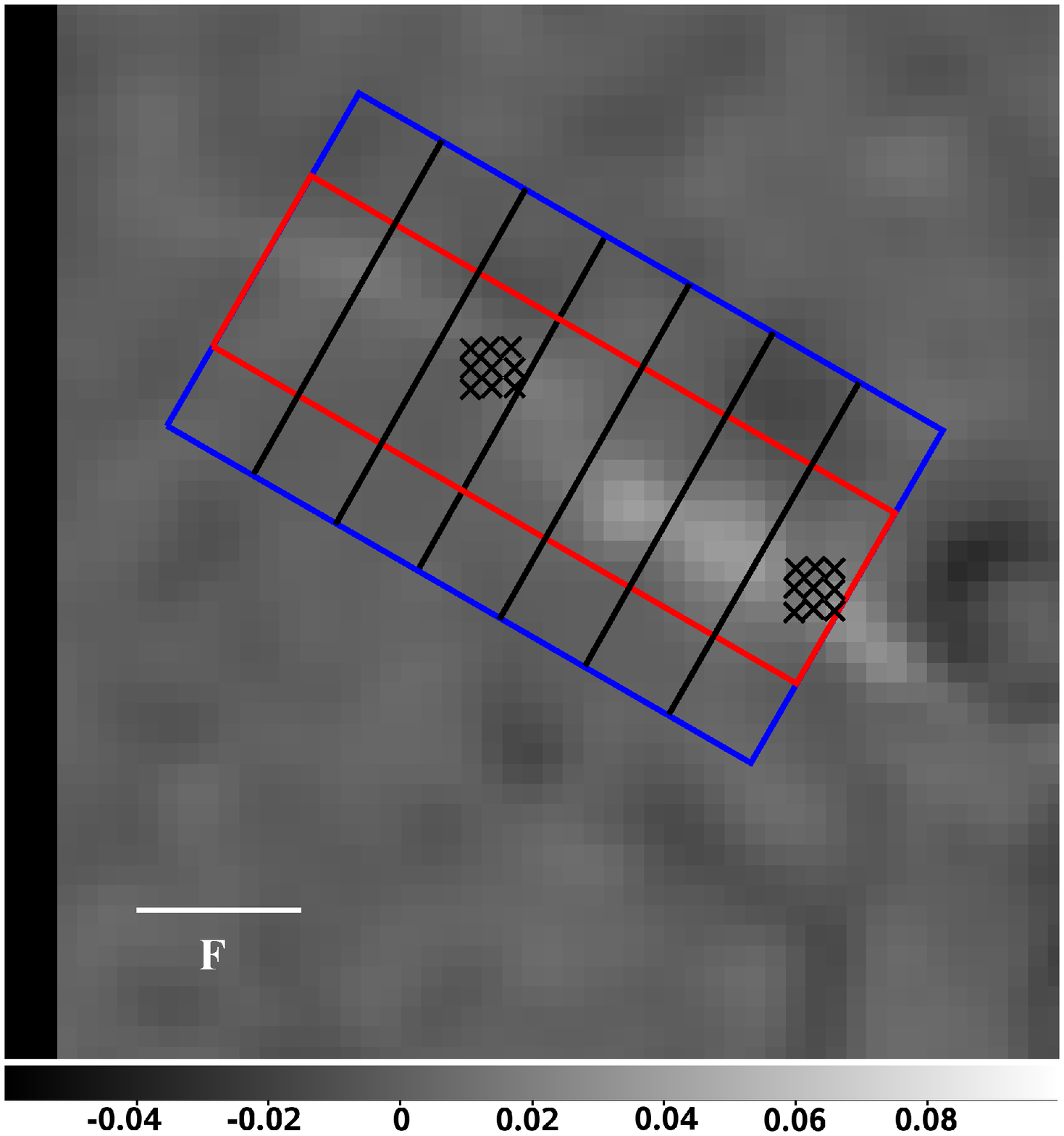} 
\\
\includegraphics[width=2.4in, angle=0]{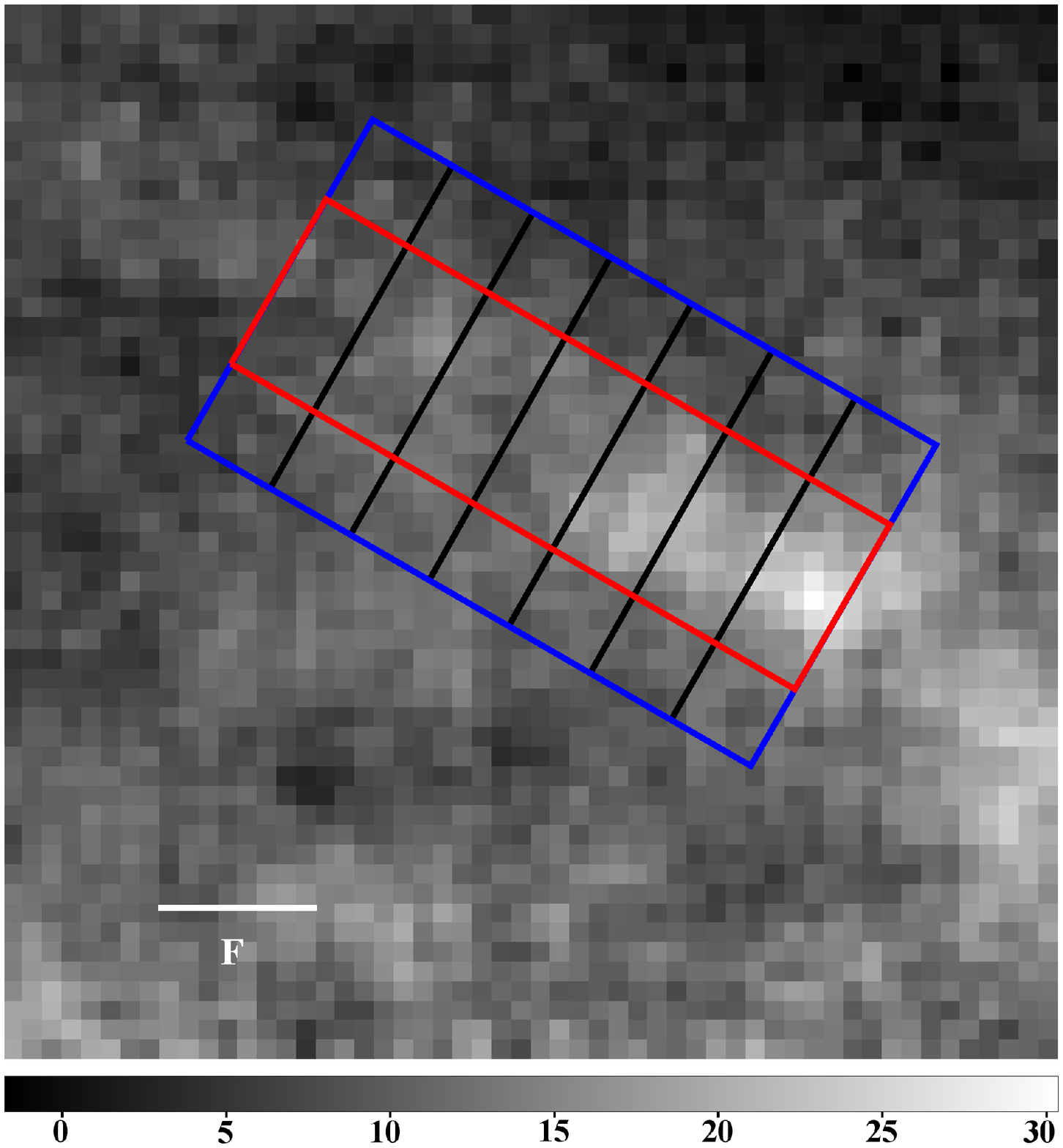}  	      
\includegraphics[width=2.4in, angle=0]{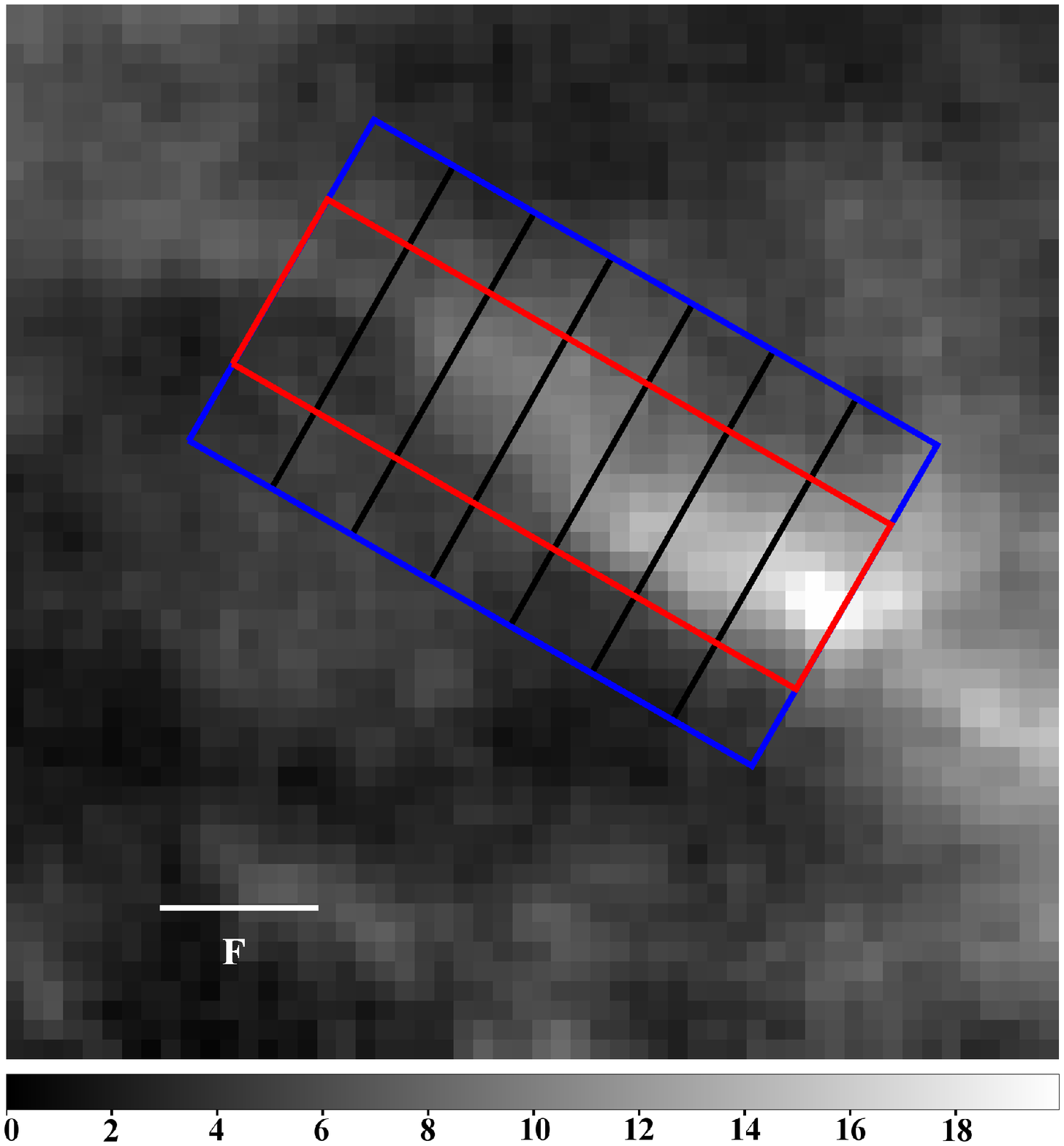}
\includegraphics[width=2.4in, angle=0]{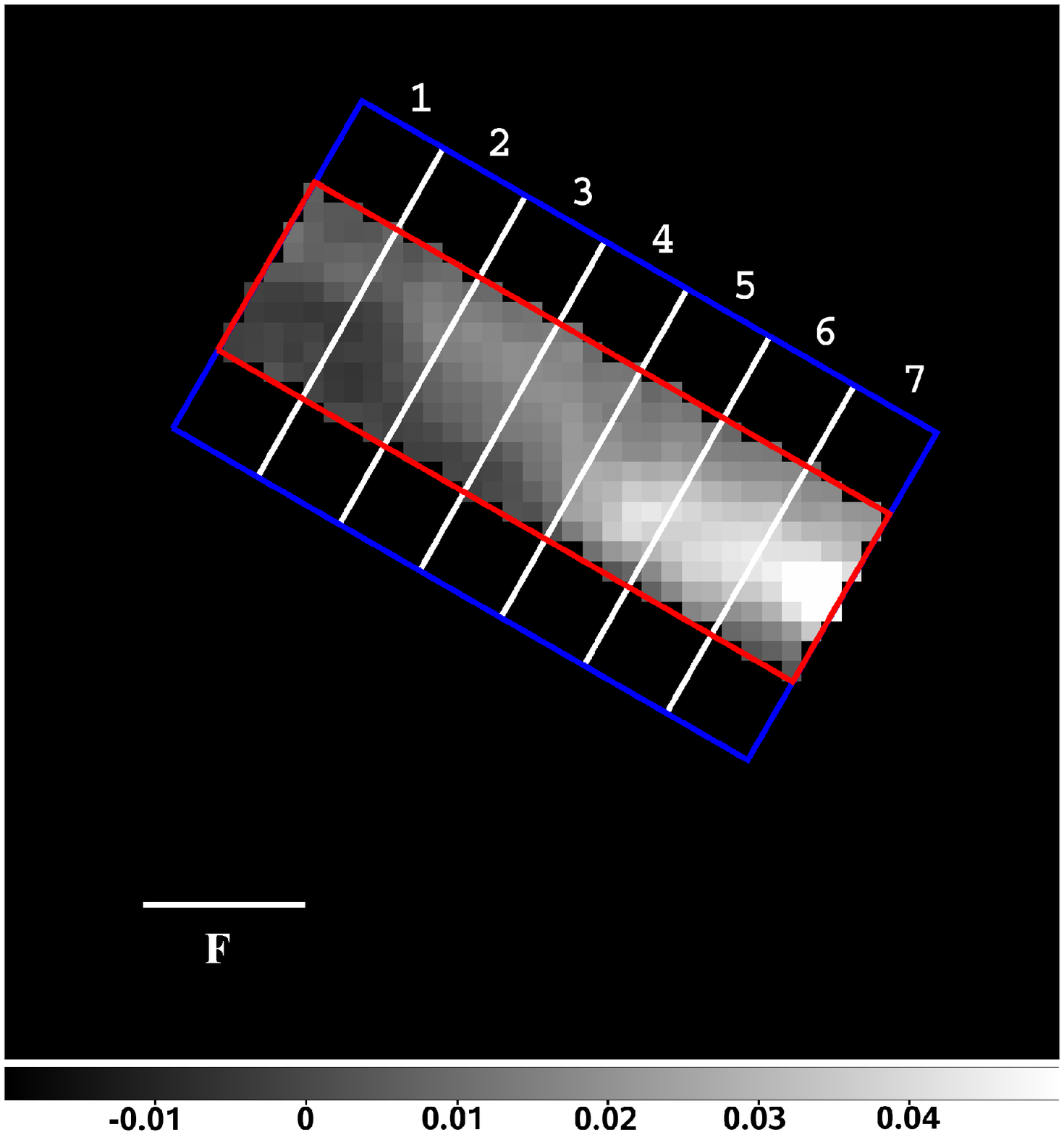}
\end{array}$
\end{center}
\caption{\footnotesize
%\small 
Morphology of IRDC F. {\it Top left:} {\it Spitzer} GLIMPSE
  IRAC 8~$\mu$m image, with linear intensity scale in MJy~$\rm
  sr^{-1}$. The horizontal line shows a scale of 3\arcmin. The image
  has 1.2\arcsec\ pixels and the PSF has a FWHM of 2\arcsec\. {\it
    Top middle:} Mass surface density, $\Sigma_{\rm SMF}$, with
  linear intensity scale in $\rm g\:cm^{-2}$, derived from the
  previous image using the small median filter MIR extinction mapping
  method of Butler \& Tan (2009). The inner, red rectangle (centered at $l=34.483^\circ$, $b=0.276^\circ$, with P.A.$=+60^\circ$ and size $0.0582^\circ$ by $0.198^\circ$) along the
  filament shows the ``on source'' region we consider to contain the
  main filamentary structure of the IRDC. The outer, blue rectangle
  extends to ``off source'' regions we consider to be representative
  of the surrounding GMC envelope. These rectangles are divided into 7
  orthogonal strips to aid in the separation of components of CO
  emission from the filament and GMC envelope. {\it Top right:} The
  same extinction map convolved with a Gaussian of 46\arcsec\ FWHM to
  match the resolution of the CO maps and pixelated to 22\arcsec on
  the same grid as the GRS survey image. The two black hatched squares
  show regions with unreliable measures of $\Sigma_{\rm SMF}$ because
  of the presence of bright MIR sources. These are excluded from the
  comparison with $\Sigco$. {\it Bottom left:} Integrated intensity map
  of $\thco$(1-0) emission over the full velocity range of $-5 - 135
  \kms$ of the GRS survey, with linear intensity scale in $\rm
  K\:km\:s^{-1}$. {\it Bottom middle:} Integrated intensity map of
  $\thco$ over the velocity range of $48-65 \kms$, i.e. the gas we
  believe is associated with the IRDC, with linear intensity scale in
  $\rm K\:km\:s^{-1}$. {\it Bottom right:} Mass surface density of the
  filament derived from $\thco$ emission, $\Sigco$, with linear
  intensity scale in $\rm g\:cm^{-2}$.}
\label{8panelF}
\end{figure}

\begin{figure}[!tb]

%\includegraphics[width=2.4in, angle=0]{f4TL.ps}f4a.ps
%\includegraphics[width=2.4in, angle=0]{f4ML.ps}f4b.ps
%\includegraphics[width=2.4in, angle=0]{f4BL.ps}f4c.ps
%\\
%\includegraphics[width=2.4in, angle=0]{f4TR.ps}f4d.ps
%\includegraphics[width=2.4in, angle=0]{f4MR.ps}f4e.ps
%\includegraphics[width=2.4in, angle=0]{f4BR.ps}f4f.ps

\begin{center}$
\begin{array}{cc}
\includegraphics[width=2.4in, angle=0]{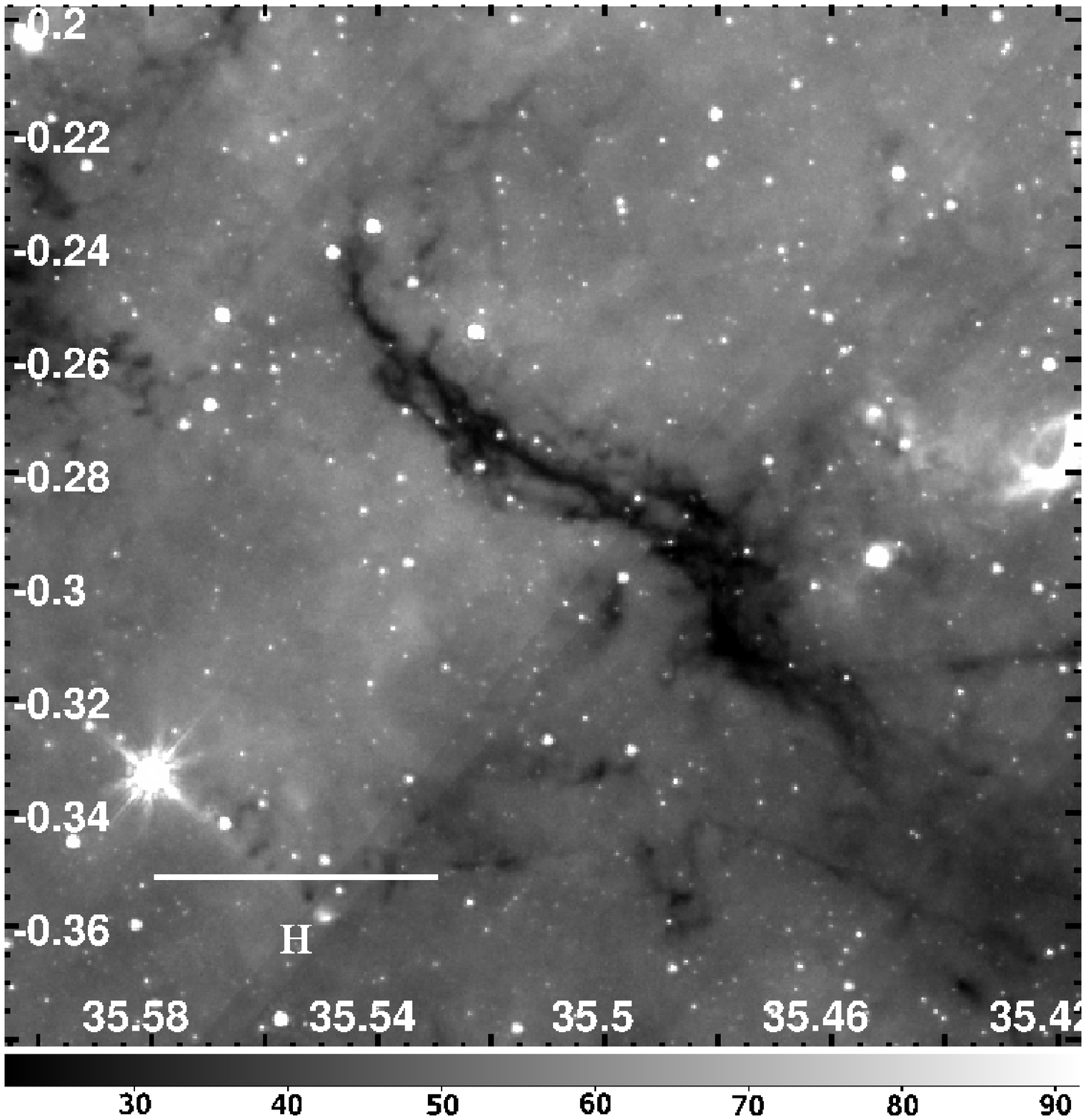}
\includegraphics[width=2.4in, angle=0]{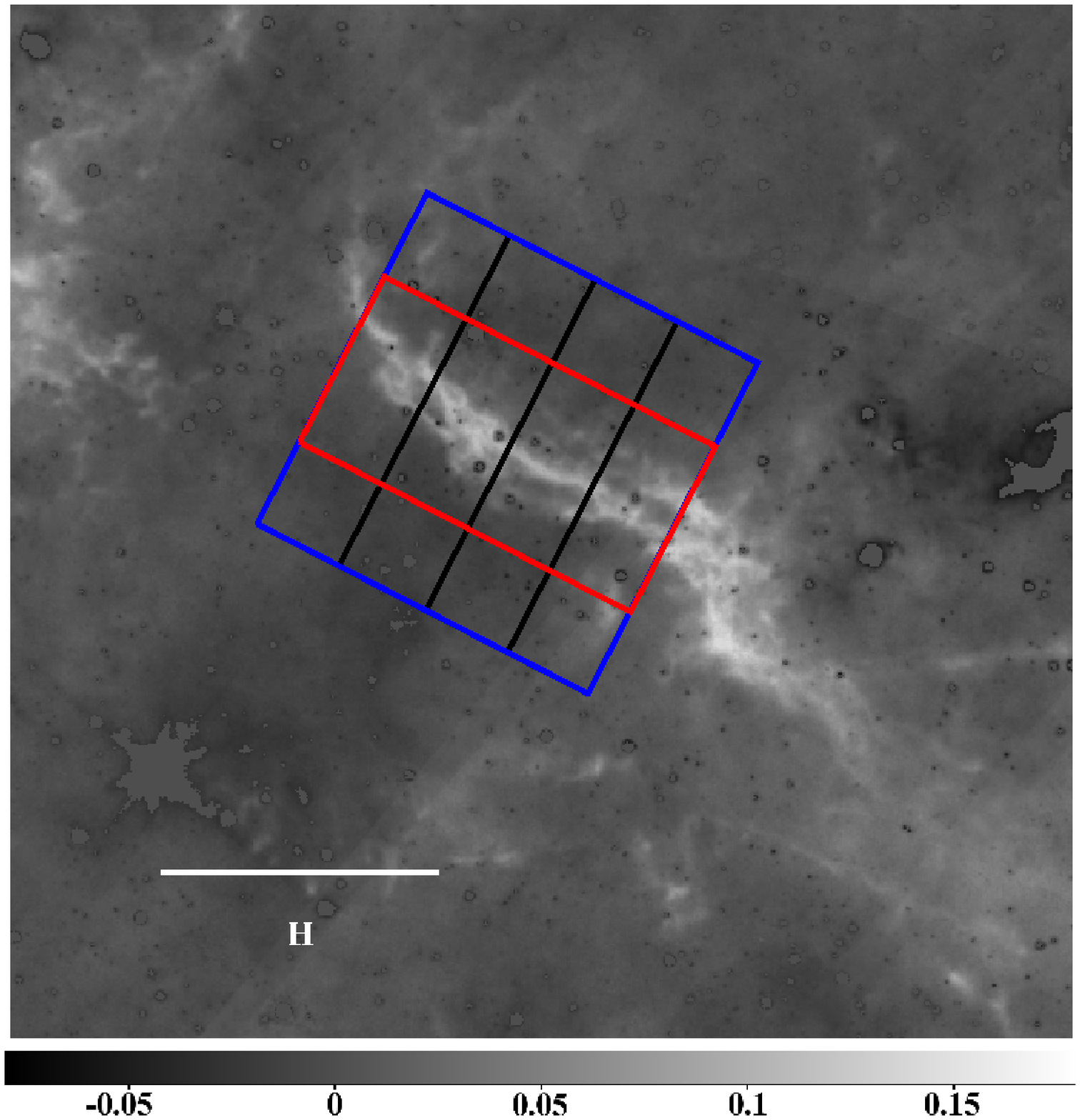}
\includegraphics[width=2.4in, angle=0]{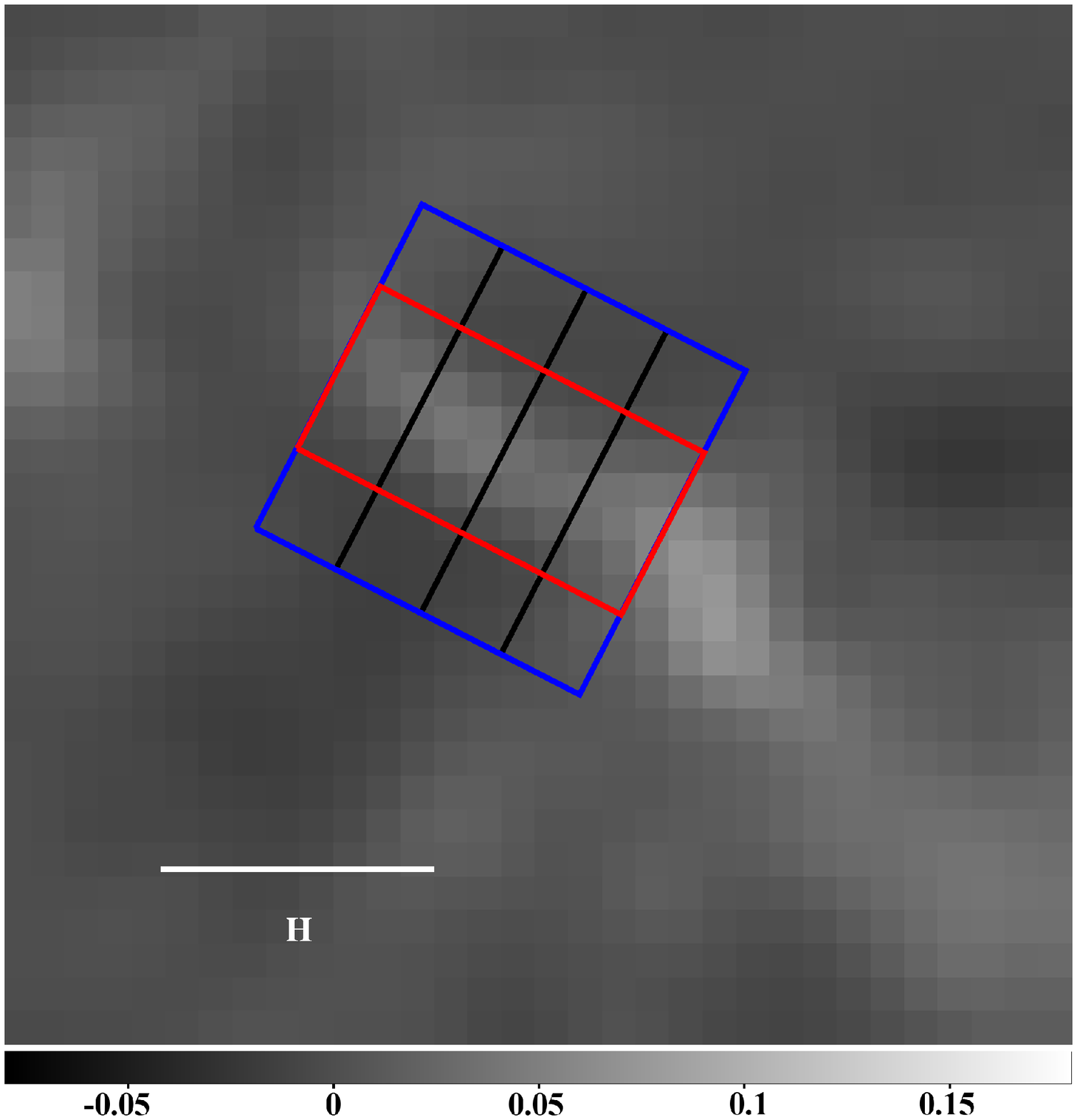}
\\
\includegraphics[width=2.4in, angle=0]{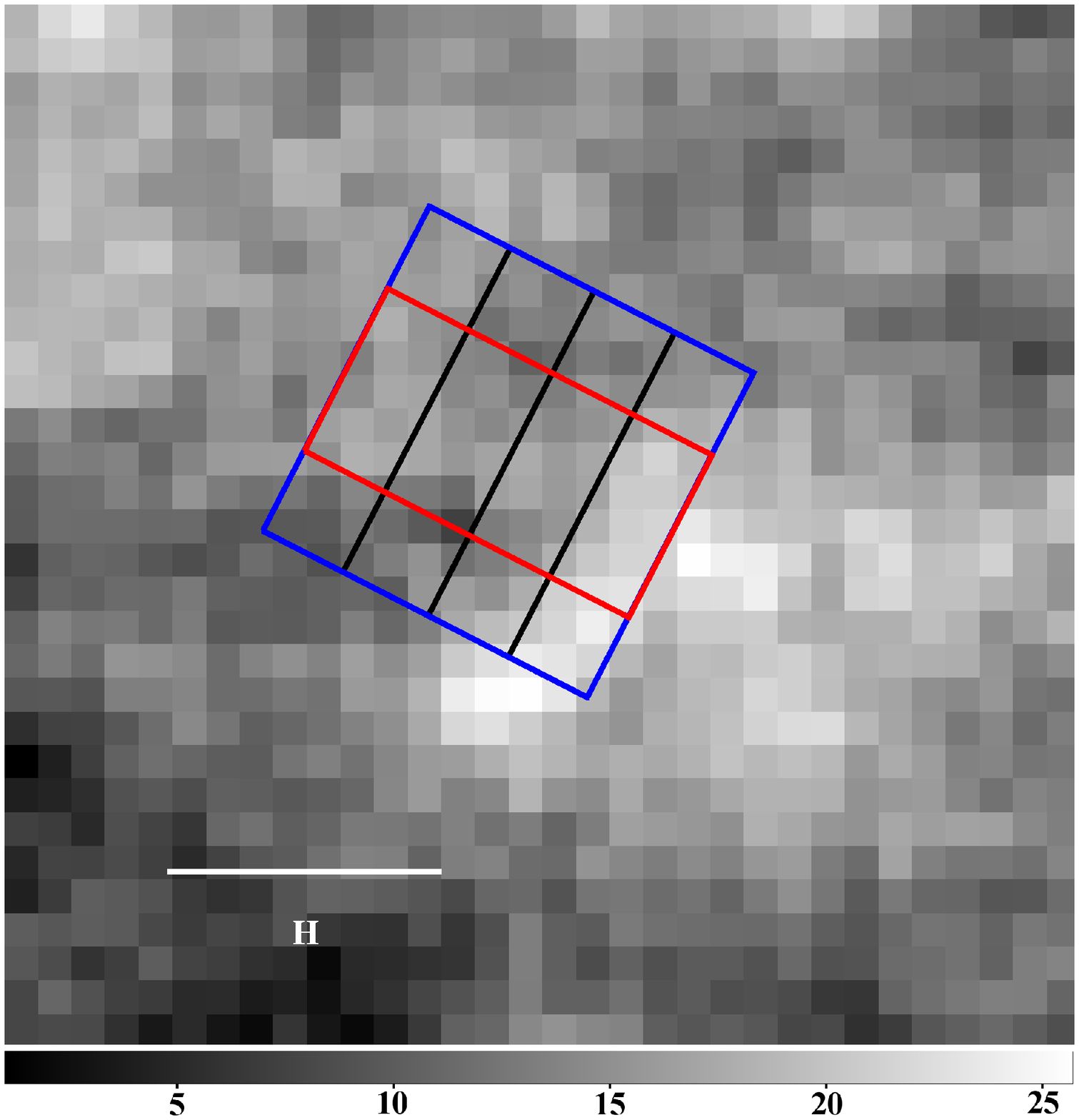}
\includegraphics[width=2.4in, angle=0]{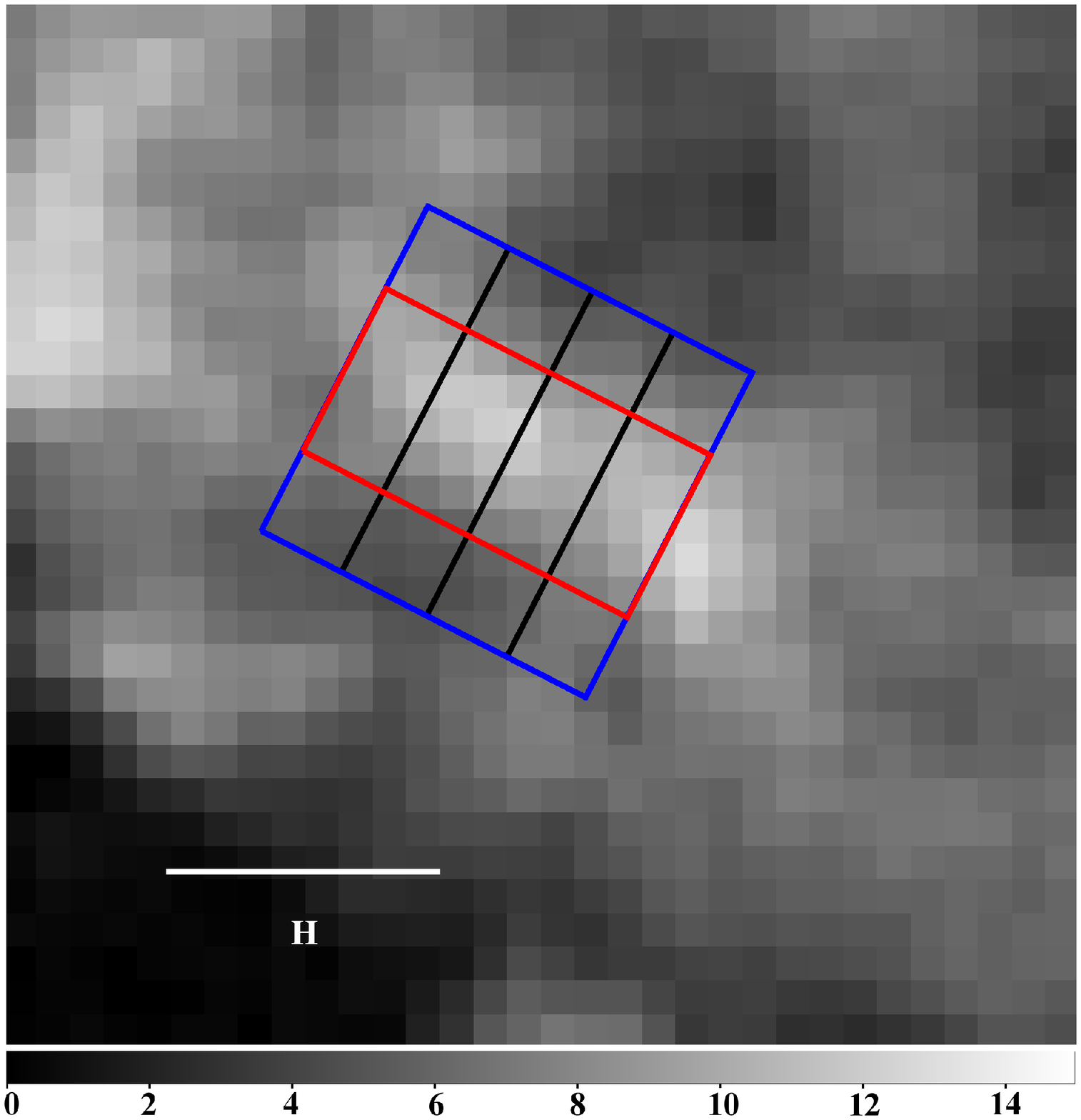}
\includegraphics[width=2.4in, angle=0]{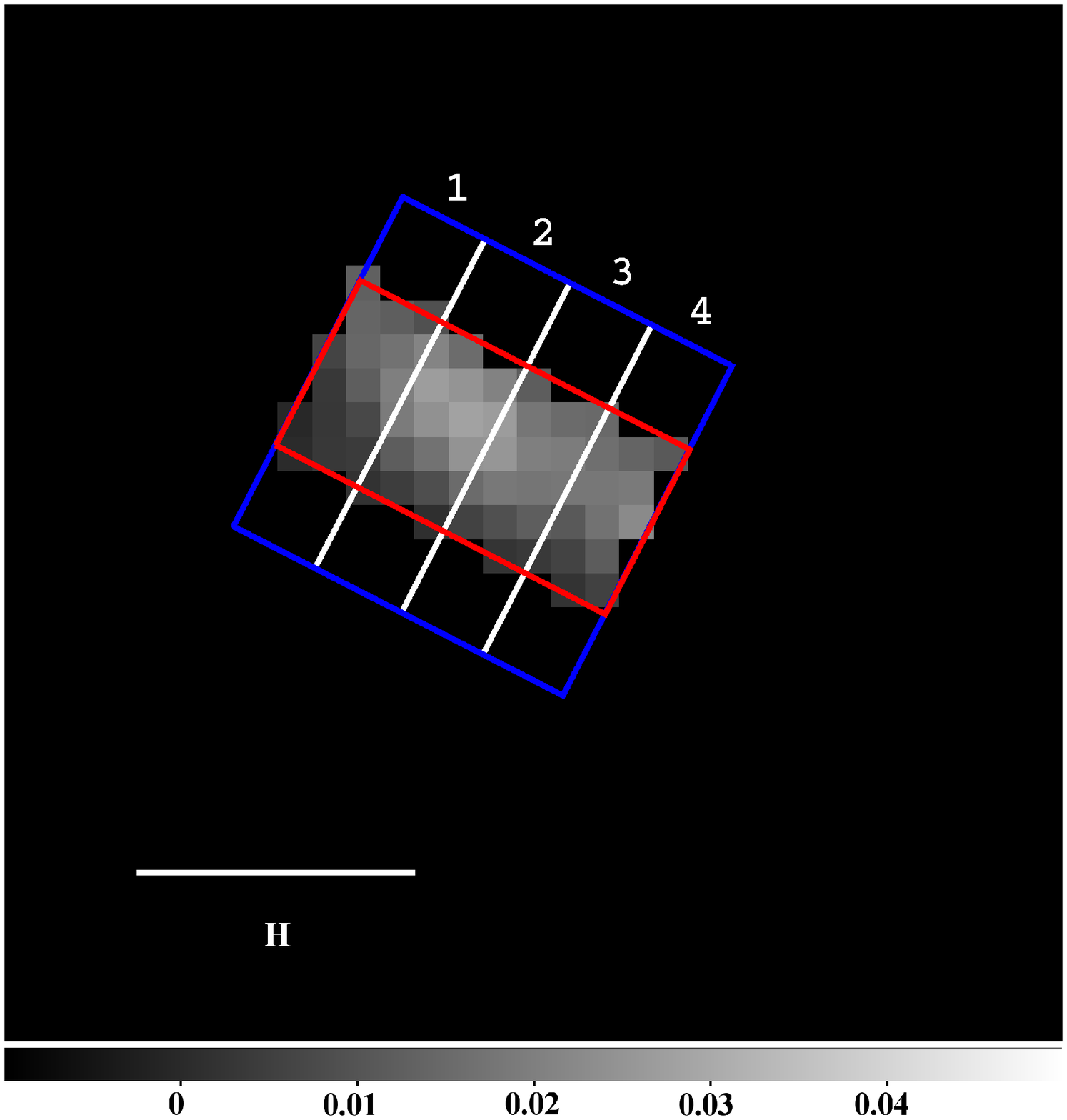}
\end{array}$
\end{center}
\caption{\small Morphology of IRDC H. {\it Top left:} {\it Spitzer} GLIMPSE
  IRAC 8~$\mu$m image, with linear intensity scale in MJy~$\rm
  sr^{-1}$. The horizontal line shows a scale of 3\arcmin. The image
  has 1.2\arcsec\ pixels and the PSF has a FWHM of 2\arcsec\. {\it
    Top middle:} Mass surface density, $\Sigma_{\rm SMF}$, with
  linear intensity scale in $\rm g\:cm^{-2}$, derived from the
  previous image using the small median filter MIR extinction mapping
  method of Butler \& Tan (2009). The inner, red rectangle (centered at $l=35.517^\circ$, $b=-0.275^\circ$, with P.A.$=+62.84^\circ$ and size $0.0307^\circ$ by $0.0637^\circ$) along the
  filament shows the ``on source'' region we consider to contain the
  main filamentary structure of the IRDC. The outer, blue rectangle
  extends to ``off source'' regions we consider to be representative
  of the surrounding GMC envelope. These rectangles are divided into 4
  orthogonal strips to aid in the separation of components of CO
  emission from the filament and GMC envelope. {\it Top right:} The
  same extinction map convolved with a Gaussian of 46\arcsec\ FWHM to
  match the resolution of the CO maps and pixelated to 22\arcsec on
  the same grid as the GRS survey image. {\it Bottom left:} Integrated
  intensity map of $\thco$ over the full velocity range of $-5 - 135
  \kms$ of the GRS survey, with linear intensity scale in $\rm
  K\:km\:s^{-1}$. {\it Bottom middle:} Integrated intensity map of
  $\thco$ over the velocity range of $40-50 \kms$, i.e. the gas we
  believe is associated with the IRDC, with linear intensity scale in
  $\rm K\:km\:s^{-1}$. {\it Bottom right:} Mass surface density of the
  filament derived from $\thco$ emission, $\Sigco$, with linear
  intensity scale in $\rm g\:cm^{-2}$.}
\label{8panelH}
\end{figure}

The GRS has an angular resolution of 46\arcsec, with sampling every
22\arcsec. The velocity resolution is 0.22~$\rm km\:s^{-1}$ (Jackson
et al. 2006).
%We refer to each (l-b-v) GRS pixel as a ``cell'' and each (l-b) GRS
%pixel simply as a ``pixel".  
From a morphological examination of the $\thco$ emission in $l,b,v$
space and comparison to the MIR extinction maps of BT09 we identify
the velocity range of the gas associated with each IRDC filament (see
Figs. \ref{spectra}, \ref{8panelF} and \ref{8panelH}). For cloud F we
consider associated gas to be at LSR velocities $48-65 \kms$ and for
cloud H at $40-50 \kms$. The total $\thco$ column is then evaluated
over the velocity range of the cloud $N_{\rm 13CO} = \int dN_{\rm
  13CO}$.

To convert from $N_{\rm 13CO}$ to total mass surface density $\Sigco$
we first assume \beq \frac{n_{\rm 12CO}}{n_{\rm 13CO}} = 6.2
\frac{R_{\rm gal}}{\rm kpc} + 18.7 \eeq (Milam et al. 2005), where
$R_{\rm gal}$ is the galactocentric radius. For clouds F and H we
estimate $R_{\rm gal} = 5.37, 5.89$~kpc, respectively (assuming
$R_{\rm gal,0}=8.0$~kpc), which would yield $n_{\rm 12CO}/n_{\rm 13CO}
= 52,55$, respectively. For simplicity we adopt $n_{\rm 12CO}/n_{\rm
  13CO} = 54$ for both clouds. We then assume \beq \frac{n_{\rm
    12CO}}{n_{\rm H2}} = 2.0 \times 10^{-4}, \eeq similar to the
results of Lacy et al. (1994). The observed variation in this
abundance in GMCs is about a factor of two (Pineda, Caselli, \&
Goodman 2008), and this uncertainty is a major contributor to the overall systematic uncertainty in our estimate of $\Sigco$. Thus our assumed abundance of $\rm ^{13}CO$ to $\rm
H_2$ is $3.70\times 10^{-6}$ and
\beq
\Sigco = 1.26 \times 10^{-2} \frac{N_{\rm 13CO}}{10^{16}{\rm cm^{-2}}}   {\rm g\: cm^{-2}},
\eeq
assuming a mass per H nucleus of $\mu_{\rm H}=2.34\times 10^{-24}\:{\rm g}$.

\begin{deluxetable}{ccccccc}
%emulate
%\rotate
%\begin{deluxetable*}{lccccc}
\tabletypesize{\footnotesize}
%\tablecolumns{6}
\tablewidth{0pt}
\tablecaption{Escape Probabilities and Critical Densities}
\tablehead{\colhead{IRDC Spectrum} &
           \colhead{$N_{\rm 13CO}$} &
           \colhead{$\Sigma_{\rm 13CO}$} &  
           \colhead{$\beta$} &
           \colhead{$n_{\rm H2}$} &  
           \colhead{$\beta n_{\rm crit}$} &  
           \colhead{$n_{\rm H2}/ (\beta n_{\rm crit})$} \\
	   \colhead{(see Fig.~\ref{spectra})} &
	   \colhead{($\rm 10^{16}\:cm^{-2}$)} &
	   \colhead{($\rm g\:cm^{-2}$)} &
	   \colhead{} &
	   \colhead{($\rm cm^{-3}$)} &
	   \colhead{($\rm cm^{-3}$)} &
	   \colhead{}
}
\startdata
F$_{\rm mean}$ & 2.73 & 0.0344 & 0.873 & 1340 & 1660 & 0.807 \\
F$_{\rm high}$ & 4.94 & 0.0622 & 0.705 & 2430 & 1340 & 1.81 \\
%F$_{\rm high}$ & 8.41 & 0.106 & 0.437 & 4200 & 830 & 5.06 \\
H$_{\rm mean}$ & 2.75 & 0.0347 & 0.696 & 2850 & 1322 &  2.15\\
H$_{\rm high}$ & 3.46 & 0.0436 & 0.587 & 3590 & 1120 & 3.22	
\enddata
\label{crittable}
\end{deluxetable}

The assumption of LTE breaks down if the density of the gas is lower
than the effective critical density, $\beta n_{\rm crit}$, i.e. the
critical density of $\thco$ (J=1-0), $n_{\rm crit}= 1.9\times
10^3\:{\rm cm^{-3}}$, allowing for radiative trapping with escape
probability $\beta=e^{-\bar{\tau}}$, where we approximate $\bar{\tau}$
as the column density weighted mean value of $\tau_\nu$. Note, for a
spherical cloud we would set $\bar{\tau}$ equal to the average of
$\tau_\nu/2$, but we use $\tau_\nu$ for these clouds with a
filamentary geometry since we expect $\tau_\nu$ seen along our line of
sight to be relatively small compared to other viewing angles. Heyer
et al. (2009) argued sub-thermal excitation of $\thco$ rotational
levels may be common in GMCs, causing lower values of line emission
than expected under LTE conditions and thus causing estimates of
$\Sigma$ based on LTE assumptions to be underestimates. The IRDC
filaments we are considering are generally of higher density than the
typical GMC volumes considered by Heyer et al. (2009). For the average
and highest intensity spectra in each IRDC (see Fig.~\ref{spectra}),
we evaluate $\beta$ (see Table \ref{crittable}). We also estimate
$n_{\rm H2}$, assuming a filament line-of-sight thickness that
is equal to its observed width. A comparison of $n_{\rm H2}$ with
$\beta n_{\rm crit}$ shows that the densities are close to or greater
than the effective critical densities, thus justifying the assumption
of LTE conditions. Furthermore, we expect that the typical density at
a given location is greater than our estimated values due to clumping
on angular scales that are smaller than the 46\arcsec\ resolution of
the CO observations. Such clumping is apparent in the MIR extinction
maps (Figs. \ref{8panelF} \& \ref{8panelH}).

%\begin{array}{cc}
%\includegraphics[width=2.7in, angle=0]{glm8F_fig_fin.ps} 
%\includegraphics[width=2.7in, angle=0]{filf_full_mom0_fin.ps}  	      
%\\
%\includegraphics[width=2.7in, angle=0]{anewsigmaafc4mapF_fig_fin.ps} 
%\includegraphics[width=2.7in, angle=0]{filf_48-65_mom0_fin.ps} 	      
%\\
%\includegraphics[width=2.7in, angle=0]{sigF_jun09_fin.ps} 
%\includegraphics[width=2.7in, angle=0]{GRSsigmamapF_REDO09_fin.ps} 

%\includegraphics[width=2.7in, angle=0]{glm8H_fig_fin.ps} 
%\includegraphics[width=2.7in, angle=0]{filh_full_mom0_fin.ps}  	      
%\\
%\includegraphics[width=2.7in, angle=0]{filh_full_mom0_fin.ps}  	      
%\includegraphics[width=2.7in, angle=0]{anewisgmaac4mapH_fig_fin.ps} 
%\includegraphics[width=2.7in, angle=0]{filh_40-50_mom0_fin.ps} 	      
%\\
%\includegraphics[width=2.7in, angle=0]{sigH_jun09_fin.ps} 
%\includegraphics[width=2.7in, angle=0]{GRSsigmamapH_REDO09_fin.ps} 

\begin{figure*}%[!tb]

\begin{center}$
\begin{array}{cc}
\includegraphics[width=3.2in,angle=0]{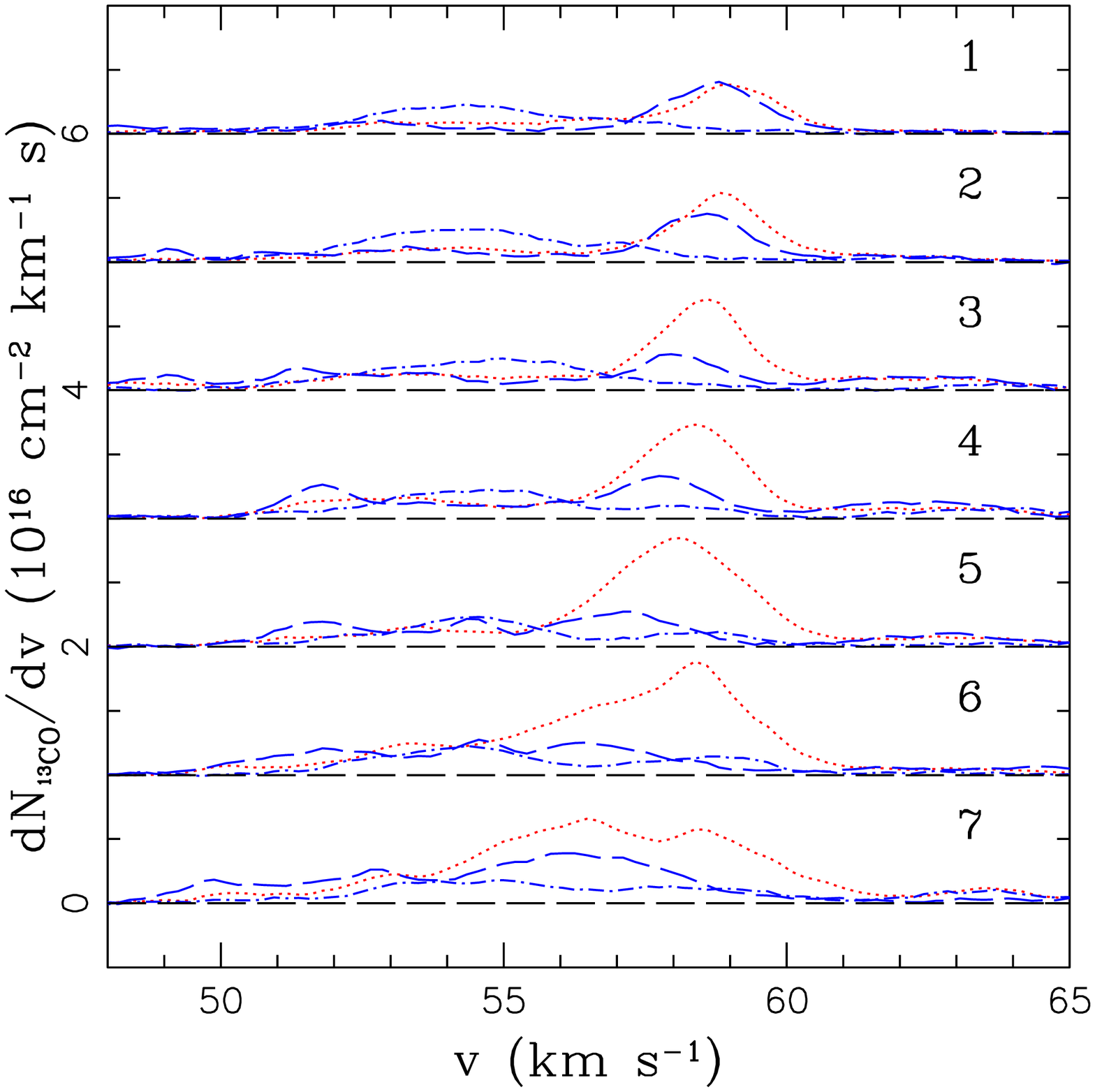} &
\includegraphics[width=3.2in,angle=0]{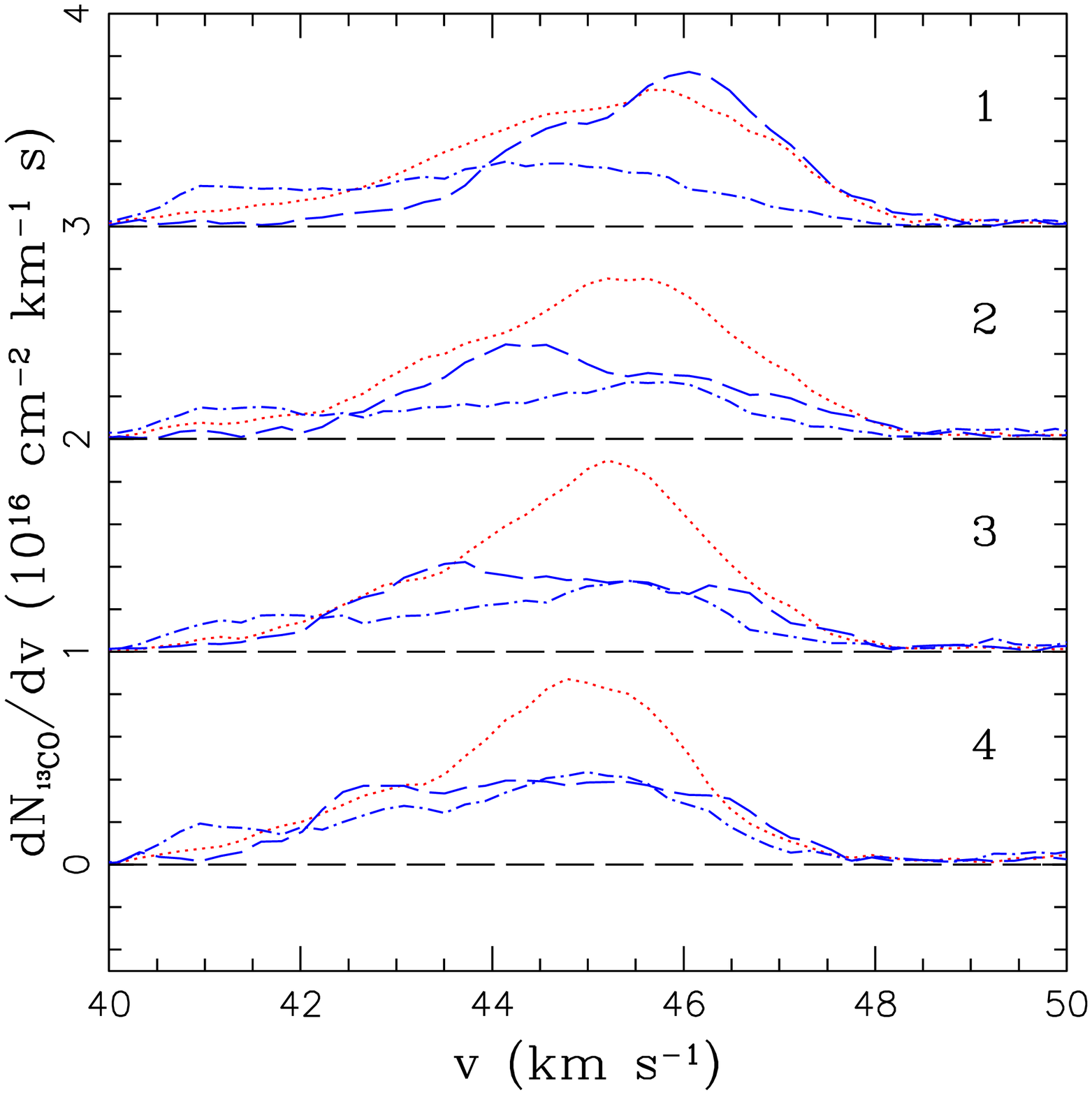} \\
\includegraphics[width=3.2in,angle=0]{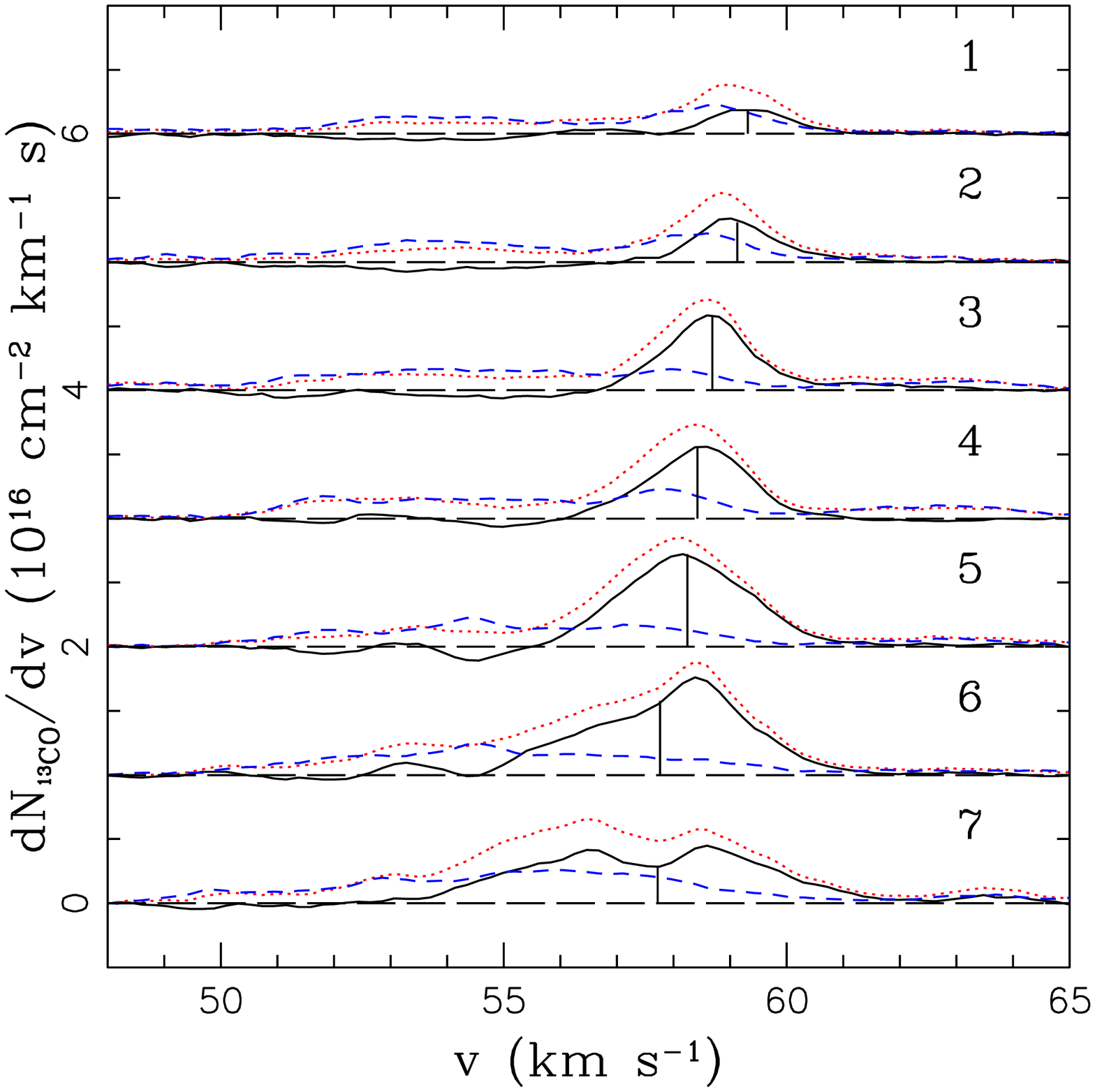} &
\includegraphics[width=3.2in,angle=0]{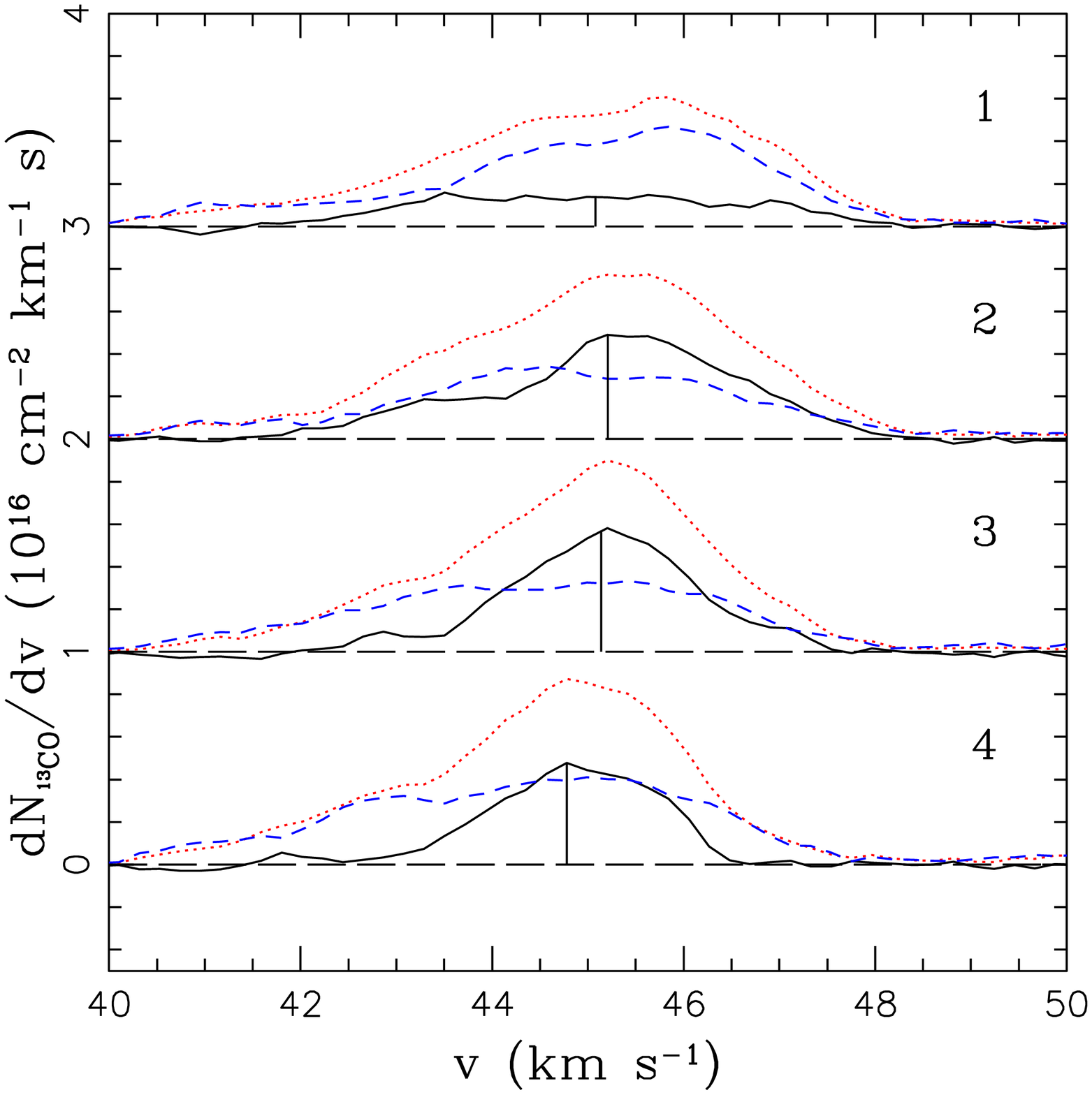} 
\end{array}$
\end{center}
\caption{
%\small 
%emulate
\footnotesize
Subtraction of IRDC envelopes. {\it (a) Top left:} Filament
  and envelope of IRDC F, with the 7 sets of spectra (top to bottom)
  corresponding to the 7 orthogonal strips (top left to bottom right)
  shown in Figure~\ref{8panelF}. In each, the dotted, red line shows
  the total $\thco$ column density distribution, including optical
  depth corrections, from the filament region, which includes an
  assumed contribution from ``envelope'' material along the line of
  sight. The long dashed and dot-dashed, blue lines show the total
  $\thco$ column density from the northern and southern envelope
  regions, respectively. {\it (b) Top right:} Filament and envelope of
  IRDC H, with the 4 sets of spectra (top to bottom) corresponding to
  the 4 orthogonal strips (top left to bottom right) shown in
  Figure~\ref{8panelH}. The line styles have the same meaning as in
  (a). {\it (c) Bottom left:} For the same strips as in (a), we
  subtract the average of the northern and southern envelope spectra
  (short dashed blue lines) from the filament (dotted red line), to
  leave an estimate of the material in the filament (solid, black
  line). The vertical solid line indicates the mean velocity. {\it (d) Bottom right:} Same analysis and labels as (c)
  applied to IRDC H.
%Note, the
%envelope-subtracted spectra are generally narrower than the total
%spectra. 
}
\label{envspectra}
\end{figure*}

\section{Comparison $\Sigco$ and $\Sigsmf$: possible trends in temperature, CO depletion and dust opacity with $\Sigma$}

The morphologies of IRDCs F and H are shown in Figures~\ref{8panelF}
and \ref{8panelH}, respectively. The $\thco$ emission is more extended
than the structure traced by the BT09 MIR extinction maps, which are
derived by comparing the observed $\rm 8\mu m$ intensity at the cloud
position with the expected background intensity interpolated from
nearby regions. Uncertainties in this estimation of the background
lead to a lack of sensitivity of the MIR extinction maps to mass
surface density contrasts $\Delta \Sigma \lesssim 0.013\:{\rm
  g\:cm^{-2}}$.

%The MIR extinction mapping technique
%assumes there is no material present here: it is only able to measure
%enhancements in $\Sigma$ relative to the surroundings. 

%Since the MIR extinction mapping method involves comparing the
%observed $\rm 8\mu m$ intensity at the cloud position with the
%expected intensity derived from nearby regions, 

Since there is a significant amount of molecular gas in the
surrounding ``envelope'' regions around the filaments, to compare the
ratio of $\Sigma_{\rm SMF}$ and $\Sigco$ in the filaments we must
first subtract the contribution to $\Sigco$ from the envelopes. To do
this we consider orthogonal strips across each IRDC filament and
measure $\thco$ emission on either side (see Figs.~\ref{8panelF} \&
\ref{8panelH}). The spectra of these ``off-source'', envelope regions
are corrected for optical depth effects, as described above, averaged
and then subtracted from the central ``on-source'', filament region
(see Fig.~\ref{envspectra}). We find the envelope-subtracted spectra
are generally narrower than the total. We also find that the size of
negative residuals created in the subtraction process are relatively
small. However, the fact that in some cases we see quite significant
variation in the envelope spectra from one side of the filament to the
other, suggests that this is one of the major sources of uncertainty
in measuring the molecular emission properties of these embedded
filaments. After this procedure, we are now in a position to compare
the envelope-subtracted values of $\Sigco$ with those derived from the
SMF extinction mapping method of BT09.

We also note that the MIR extinction derived values of $\Sigma$ suffer
from their own systematic uncertainties, including corrections due to
foreground dust emission (BT09), scattering in the IRAC array
(Battersby et al. 2010), adopted dust opacities and dust to gas
ratios. However, saturation due to large $\rm 8\mu m$ optical depth
and/or an insufficiently subtracted foreground is not an important
source of error for our present study, since the values of $\Sigsmf$
are in any case reduced when the maps are smoothed to the lower
resolution of the $\thco$ data.

As noted by BT09, the MIR extinction mapping technique fails for
locations where there are bright MIR sources. If the intensity of the
source is greater than the background model, then formally an
unphysical, negative value of $\Sigma$ is returned by this method. In
the analysis of BT09, negative values of $\Sigma$ are allowed up to a
certain threshold value to account for noise-like, approximately
Gaussian, fluctuations in the background intensity. For the more
extreme fluctuations caused by bright sources, BT09 set $\Sigma=0$.
%(although for certain averages these pixels are simply excluded). 
Thus the effect of a bright MIR region within an IRDC is to cause the
extinction mapping method to underestimate the true mass surface
density. For most point-like MIR sources, this effect is quite minor
after the extinction maps are averaged to the 22.14\arcsec\ pixel
scale of the CO observations. However, in IRDC F we identify two regions
(indicated in Fig.~\ref{8panelF}), which are significantly affected by
bright MIR sources and exclude them from the subsequent analysis.

\begin{figure}%[!tb]

\begin{center}$
\begin{array}{cc}
\includegraphics[width=3.2in]{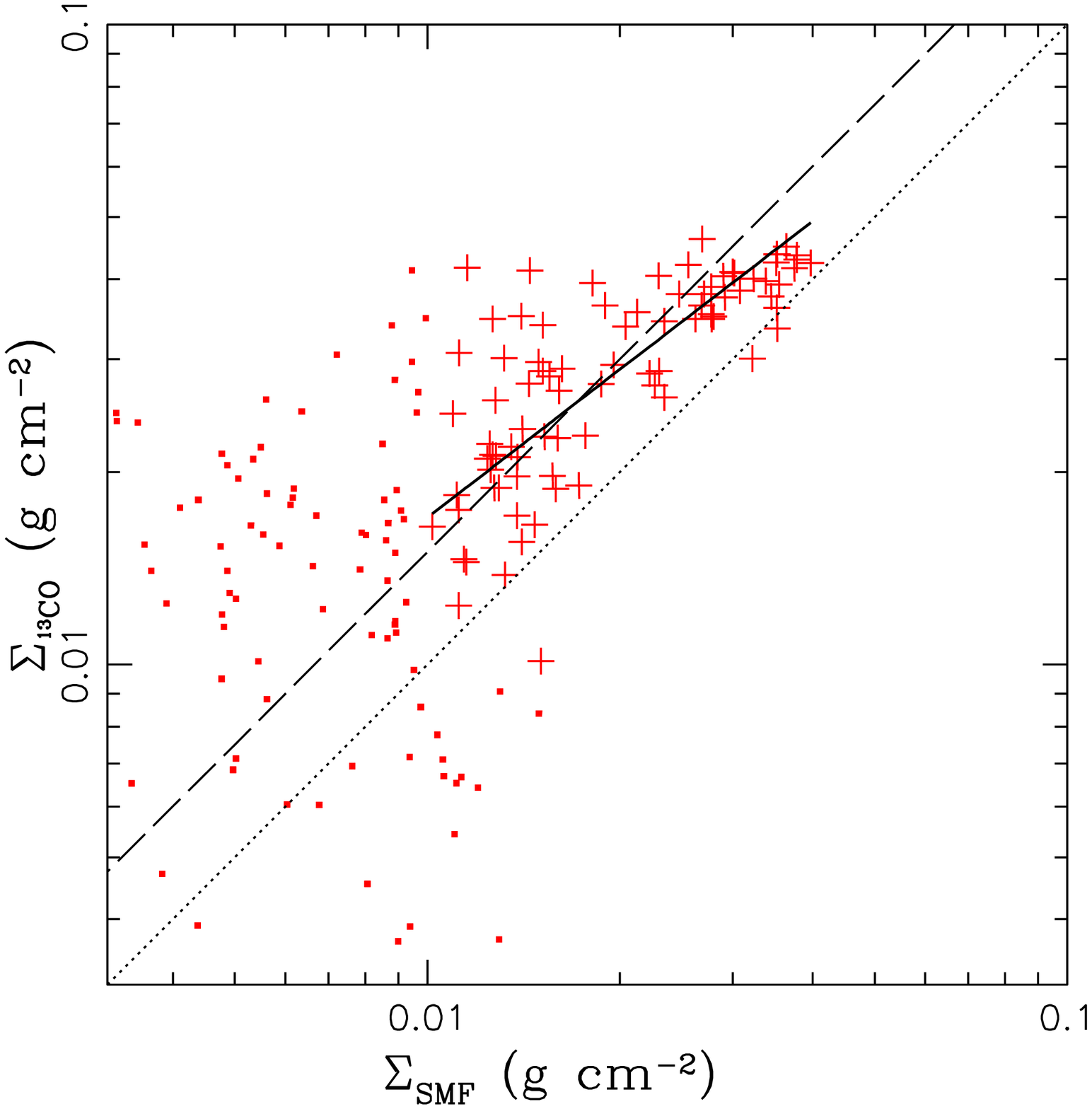} & 
\includegraphics[width=3.2in]{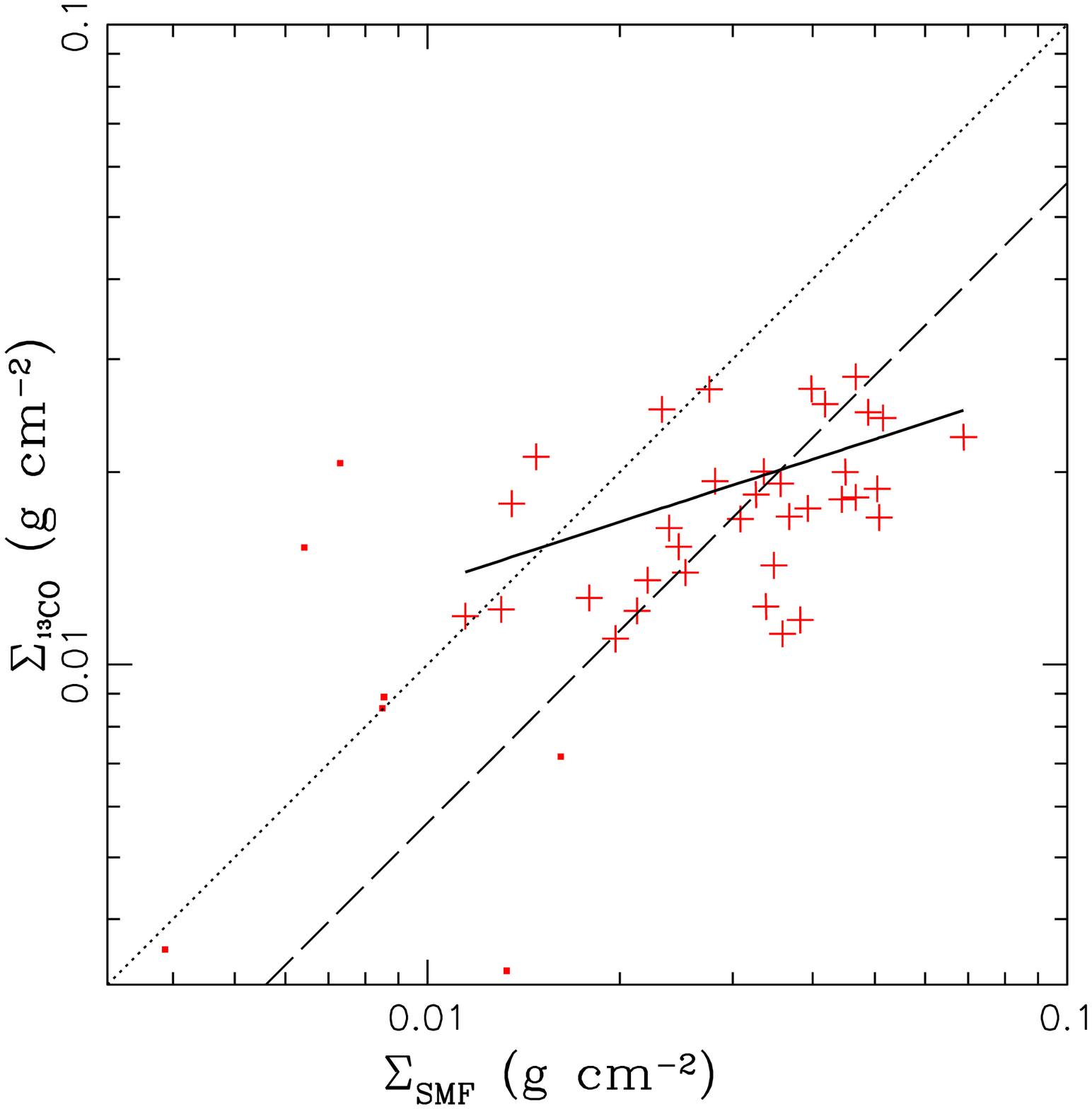}\\
\includegraphics[width=3.2in]{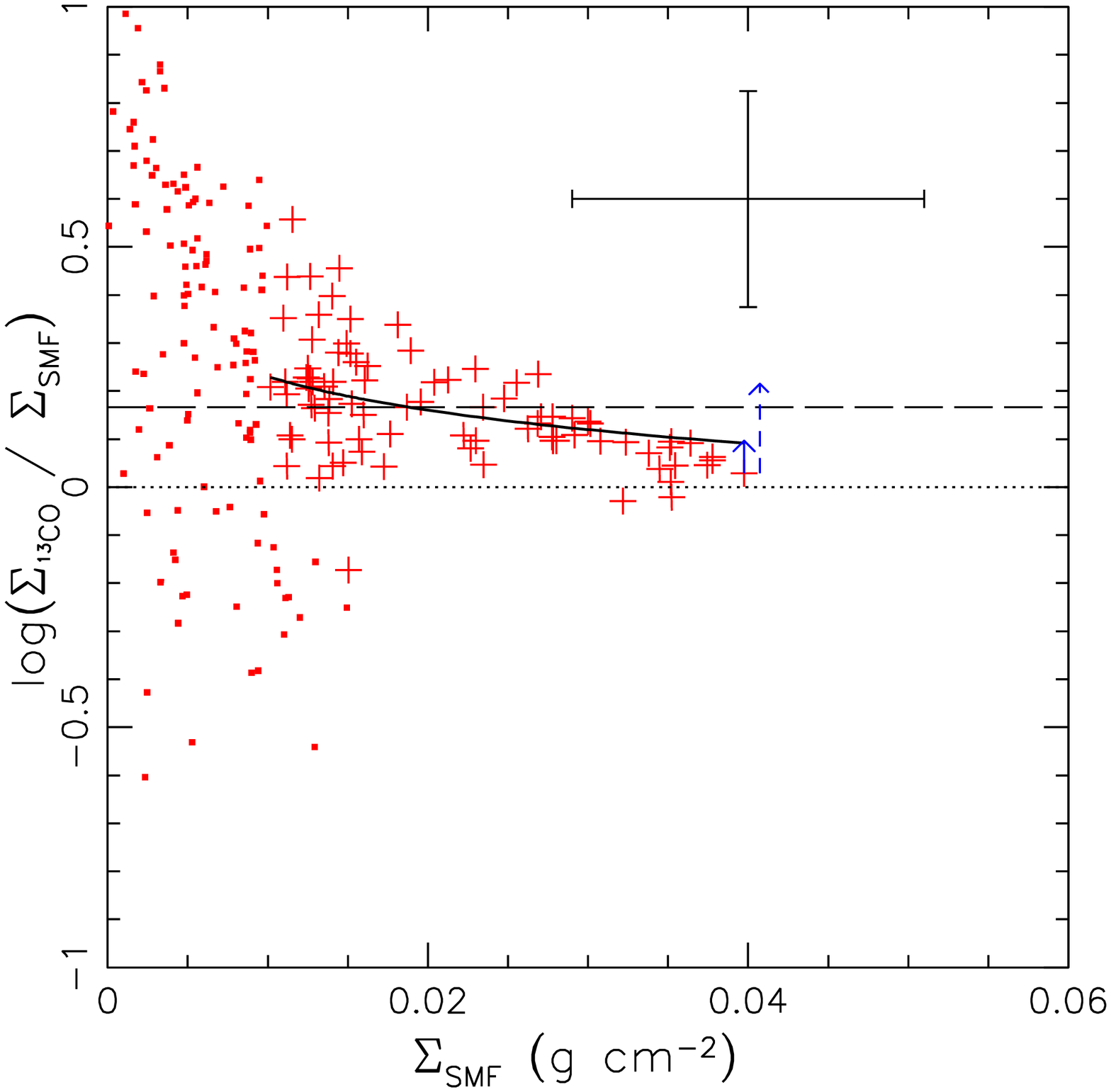} & 
\includegraphics[width=3.2in]{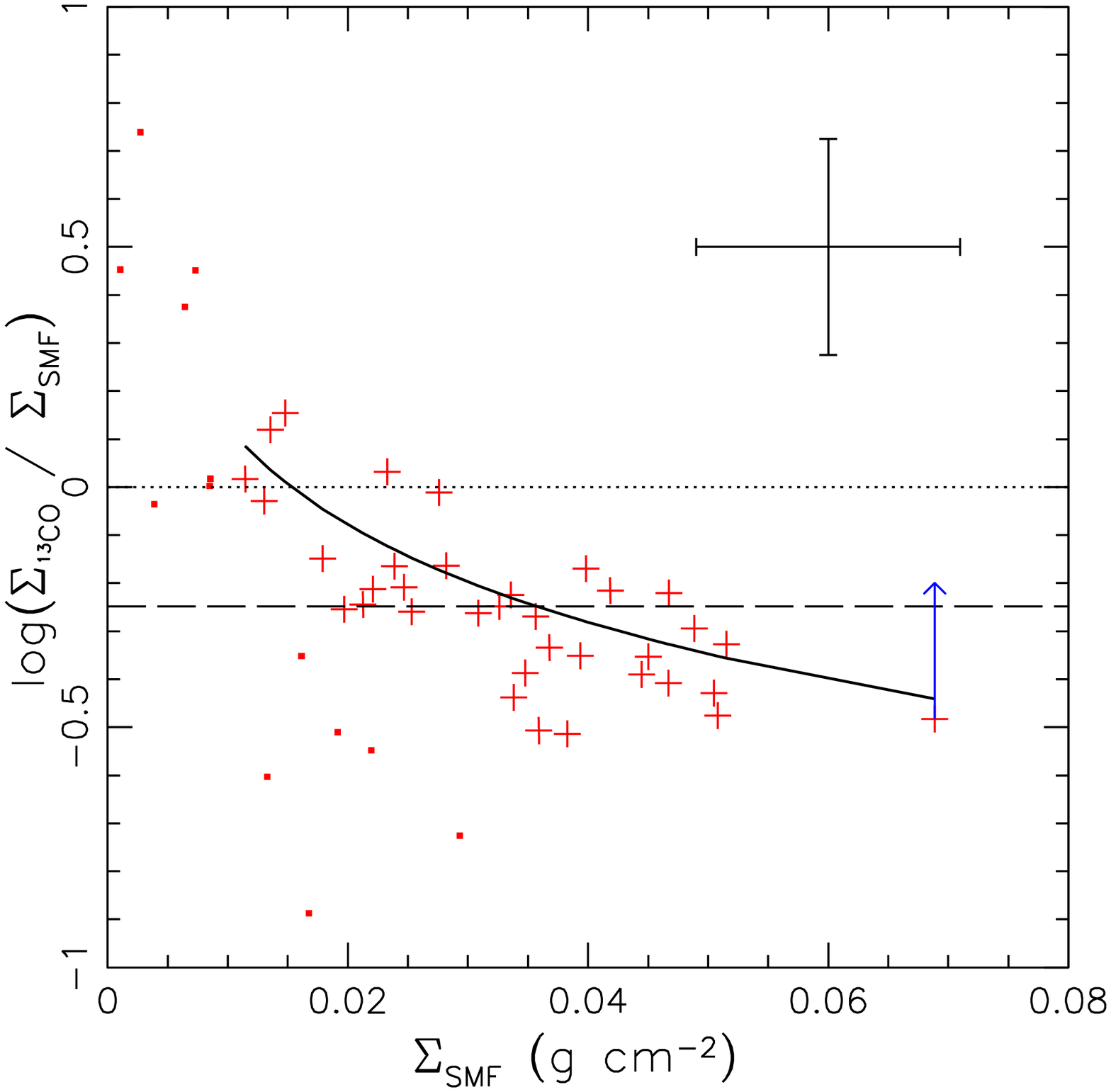}
\end{array}$
\end{center}
\caption{\small {\it (a) Top Left:} Direct comparison of $\Sigco$ (after
  envelope subtraction) and $\Sigsmf$ in IRDC F. The crosses show
  locations where both $\Sigco$ and $\Sigma_{\rm SMF}>0.01\:{\rm
    g\:cm^{-2}}$. The dots show locations of lower surface
  densities. The uncertainties in the individual measurements are
  assumed to be 15\% plus a systematic error of 0.01~$\rm
  g\:cm^{-2}$. The dotted line shows the one to one linear relation
  and the long dashed line shows the best-fit offset linear
  relation. The solid line shows the best-fit power law relation (see
  text). {\it (b) Top Right:} Same as (a) but for IRDC H. {\it (c)
    Bottom Left:} Logarithm of the ratio of $\Sigco$ to $\Sigsmf$ as a
  function of $\Sigma_{\rm SMF}$ for IRDC F, with the same symbol
  notation as in (a). The cross in the upper-right corner indicates
  typical estimated uncertainties. The solid (dashed) arrow shows the
  effect on the highest $\Sigma$ position of reducing the assumed
  temperature from 20~K to 15~K (10~K). {\it (d) Bottom Right:} Same
  as (c) but for IRDC H. The arrow shows the effect on the highest
  $\Sigma$ position of reducing the assumed temperature from 15~K to
  10~K.  }
\label{Sigcomp}
\end{figure}

The values of $\Sigco$ (after envelope subtraction) and $\Sigma_{\rm
  SMF}$ for each pixel in the filament regions of IRDCs F and H are
compared in Figure~\ref{Sigcomp}. We note that both these measures of
$\Sigma$ are independent of the distance to the cloud. Considering
just the data with $\Sigco$ and $\Sigma_{\rm SMF}>0.01\:{\rm
  g\:cm^{-2}}$, the best fit power law relation $\Sigco/{\rm
  g\:cm^{-2}} = A (\Sigma_{\rm SMF}/{\rm g\:cm^{-2}})^\alpha$ has
$\alpha=0.77\pm0.18, 0.32\pm 0.28$ and $A=0.58 \pm 0.44, 0.074 \pm
0.069$ for IRDCs F and H, respectively.

Over the range $0.01<\Sigma_{\rm SMF}/{\rm g\:cm^{-2}}<0.07$ we find
that $\Sigco \simeq \Sigsmf$ to within a factor of $\sim 2$ for the
mean values in each IRDC filament. The dispersion within an individual
IRDC from pixel to pixel is also at about this level. The systematic
offsets may reflect real systematic variations of the assumed
temperature, $\rm ^{13}CO$ abundance, dust opacities or envelope
subtraction method for each IRDC compared to our adopted values and
methods. The dispersion may reflect local systematic variations and
errors introduced by measurement noise.

In both IRDCs the ratio $\Sigco/\Sigma_{\rm SMF}$ decreases with
increasing mass surface density as measured by $\Sigsmf$. We gauge the
significance of this trend by noting that the above values of $\alpha$
for IRDCs F and H differ from unity by 1.3 and 2.4 standard
deviations, respectively, assuming errors are distributed normally.

There are several physical processes that may be causing such a
trend. A systematic temperature decrease from the lower column
density, outer regions of the filaments to the higher column density
centers would lead us to systematically underestimate $\Sigco$ in the
latter, while having little direct effect on dust opacities (although
there may be an indirect effect via formation of ice mantles, see
below). For the temperature ranges we expect to be present, i.e. from
$\sim$10-20~K (e.g. Pillai et al. 2006 - although note these are based
on measurements of $\rm NH_3$, which may trace different conditions to
those of the $\thco$), this effect becomes important for values of
$T_{B,\nu} \gsim 4$~K (see Fig. \ref{fig:tauvsT}b). We illustrate the
size of this effect for the highest $\Sigsmf$ positions in IRDCs F and
H in Fig.~\ref{Sigcomp}. A systematic temperature decrease of 5~K in
the high $\Sigma$ regions could remove much of the observed trend.

Another possibility is that our corrections for the optical depth of
the $\thco$ emission are systematically underestimated at the higher
column density positions. This would be expected if a significant
amount of mass is contained in unresolved dense cores. The largest
optical depth corrections in the highest column density positions
presently lead to an increase in the estimated column by factors
$\lsim 2$. Future higher angular resolution $\thco$ and $\rm C^{18}O$
observations of these filaments are required to investigate this issue
further.

Depletion of CO molecules onto dust grains is known to occur in cold,
high volume density gas (Caselli et al. 1999). This process could
systematically reduce the gas phase CO abundance in the high $\Sigma$
regions of the IRDCs. For depletion to be fully responsible for the
observed trends in IRDCs F and H would require about a factor of 2
depletion as $\Sigsmf$ increases from 0.01 to 0.05~$\rm g \:cm^{-2}$.
Because of depletion, CO is not expected to be an ideal tracer of the
densest, coldest parts of IRDCs. High resolution studies of other
tracers, such as $\rm NH_3$, will likely be needed to measure the
kinematics of these regions.

The depletion of CO via formation of CO ice mantles on dust grains
would also have some effect on the MIR opacities of these grains, thus
affecting our measurement of $\Sigsmf$. The grains would become larger
and absorption features due to pre-existing water ice mantles may
become obscured. The Ossenkopf \& Henning (1994) grain models show a
$\sim$50\% increase in $\rm 8\mu m$ opacity, $\kappa_{\rm 8\mu m}$,
going from bare grains to those with thick ice mantles. The BT09
estimates of $\Sigsmf$ assumed a constant value of $\kappa_{\rm 8\mu
  m}$ consistent with the thin ice mantle model of Ossenkopf \&
Henning (1994). Thus the observed decrease in the ratio of
$\Sigco/\Sigsmf$ with increasing mass surface density could be caused by
thickening of grain ice mantles, causing us to systematically
overestimate $\Sigsmf$.

Another possible explanation for a trend of decreasing
$\Sigco/\Sigsmf$ with increasing mass surface density is active chemical
fractionation (Langer et al. 1980, 1984; Glassgold, Huggins, \& Langer
1985; Visser, Van Dishoeck, \& Black 2009), which enhances the
abundance of $\thco$ in regions where $\rm ^{13}C^+$ is present via
the via the ion-molecule exchange reaction (Watson, Anicich, \&
Huntress 1976), $\rm {^{12}CO} + {^{13}C^+} \rightarrow {^{12}C^+} +
{^{13}CO} +35\:{\rm K}$. FUV irradiation maintains a relatively high abundance
of $\rm ^{13}C^+$ in the outer regions of the cloud, with $A_V\lesssim
1$~mag.
%with $N_{\rm H} \sim 10^{20} {\rm  cm}^{-2}$. 
The relative abundance of $\thco$ to $\rm ^{12}CO$ can be enhanced by
factors of $\sim 10$. For $A_V \gtrsim 2$, the enhancement factor is
only about 20\% (e.g. for a model with $n_{\rm H_2}=10^3\:{\rm
  cm^{-3}}$ and T=20~K; Glassgold et al. 1985). For IRDCs F and H, the
filaments are typically embedded inside ``envelope'' gas with
$\Sigco\simeq 0.01\:{\rm g\:cm^{-2}}$, i.e. $N_{\rm H}\simeq 4\times
10^{21}\:{\rm cm^{-2}}$. The trend of decreasing $\Sigco/\Sigsmf$ is
observed to occur as $\Sigsmf$ increases to $\sim 0.05\:{\rm
  g\:cm^{-2}}$. $\thco$ fractionation enhancements should not be
varying in this regime by large enough factors to explain our observed
results. It should also be noted that even in the lower-column density
regions, if the FUV flux is large enough, then isotope selective
photodestruction of $\thco$ compared to the self-shielded $\rm
^{12}CO$ (Bally \& Langer
1982) may reverse the fractionation-produced enhancement of the
abundance of $\thco$ to $\rm ^{12}CO$.

\section{The Dynamical State of the IRDC Filaments}

\subsection{IRDC Masses from $\thco$ Emission and MIR Dust Absorption}

We calculate IRDC masses using the observed mass surface densities and
angular sizes, and by assuming near kinematic distances, since the
clouds are seen in absorption against the Galaxy's diffuse MIR
emission. We adopt the kinematic distances of Simon et al. (2006),
who assumed the Clemens (1985) rotation curve.
%, a distance of the Sun to the Galactic center of 8.5~kpc and a local rotational velocity of 220 $\kms$. 
This leads to a distance of 3.7~kpc for IRDC F and 2.9~kpc
for IRDC H. We assume uncertainties of 0.5~kpc, which could result,
for example, from line of sight noncircular motions of $\sim8\:{\rm
  km\:s^{-1}}$. Temperature uncertainties of 5~K would lead to
$\Sigco$ uncertainties of $\sim20\%$. We estimate similar levels of
uncertainty in $\Sigsmf$ due to foreground correction and background
interpolation uncertainties (BT09). Then, for IRDC F, the
$\thco$-derived mass assuming $T=20$~K in the on-source filament
region after envelope subtraction is $M_{\rm 13CO} =
3300\pm1100\:M_\odot$ and the dust extinction mass is $M_{\rm SMF} =
1900\pm 640\: M_\odot$. For IRDC H, we find $M_{\rm 13CO} (T=15\:{\rm
  K}) = 370\pm150\:M_\odot$ and $M_{\rm SMF} = 580\pm 230\:
M_\odot$. For our calculations involving the total mass of the clouds
we take averages of the above estimates while still assuming 20\%
uncertainties in the averaged values of $\Sigma$. Thus we adopt
$M=2600\pm 870\:M_\odot$ for IRDC F and $M=480\pm 190\:M_\odot$ for
IRDC H. These results are summarized in Table~\ref{masstable1}.

%We now compare these mass estimates with those from derived from
%dynamical estimates assuming virial equilibrium.

\begin{deluxetable}{ccc}
%emulate
%\rotate
%\begin{deluxetable*}{lccccc}
\tabletypesize{\footnotesize}
%\tablecolumns{5}
\tablewidth{0pt}
\tablecaption{IRDC Ellipsoidal Virial Analysis}
\tablehead{\colhead{Cloud property} &
           \colhead{IRDC F} &
           \colhead{IRDC H} 
}
\startdata
d (kpc) & $3.7\pm0.5$ & $2.9\pm0.5$\\
R (pc) & $1.88\pm0.25$ & $0.91\pm0.16$\\
$R_{\rm obs}$ (pc) & $3.52\pm 0.48$ & $1.21\pm 0.21$\\
$y\equiv Z/R$ & $3.93\pm 0.6$ & $2.06 \pm 0.3$\\
$\Sigma_{\rm 13CO}$ ($\rm g\:cm^{-2}$) & $0.0147\pm0.003$ & $0.0133\pm0.003$ \\
% & 0.0147 & 0.01328\\
$\Sigma_{\rm SMF}$ ($\rm g\:cm^{-2}$)& $0.0082\pm0.0016$ & $0.0209\pm0.004$\\
% & 0.008229 & 0.02087\\
$\Sigma_{\rm 13CO}({\rm env})$ ($\rm g\:cm^{-2}$) & $0.0196\pm0.004$ & $0.0212\pm0.004$\\
$M_{\rm 13CO}$ ($M_\odot$) & $3300\pm1100$ & $370\pm150$\\
$M_{\rm SMF}$ ($M_\odot$)& $1900\pm640$ & $580\pm230$\\
$M$ ($M_\odot$)& $2600\pm 870$ & $480 \pm 190$\\
$a_1$ & 10/9 & 10/9\\
$a_2$ & 0.538 & 0.750\\
$W$ ($10^{46}$erg)& $-11.0\pm7.1$& $-1.08\pm0.78$\\
$\sigma$ (km/s) & $1.46\pm0.15$ & $1.20\pm0.12$\\
%$\sigma$ (km/s) & 1.27 & 1.13\\old submitted
$t_s=2R/\sigma$ (Myr) & $2.5\pm0.4$ & $1.5\pm0.3$ \\
%$t_s=2R/\sigma$ (Myr) & $2.9\pm0.37$ & $1.6\pm0.27$ \\old submitted
${\cal T}$ ($10^{46}$erg) & $16.4\pm6.4$ & $2.06\pm0.91$\\
${\cal T}_0(A)$ ($10^{46}$erg) & $14\pm8$ & $2.7\pm1.8$\\
${\cal T}_0(B)$ ($10^{46}$erg) & $73\pm30$ & $4.3\pm2.0$\\
$\alpha\equiv 5\sigma^2R/(GM)$ & $1.78\pm 0.73$ & $3.17\pm 1.50$\\
\enddata
\label{masstable1}
\end{deluxetable}

%F & $3.7\pm0.5$ & $3300\pm900$ & $1900\pm500$ & $2600\pm 700$ & $4.0\pm 0.6$ & $1.88\pm0.25$ & $-10.8\pm6.0$ & 1.27 & 1.45 & $12.5\pm3.4$ & $14.5\pm5.0$\\
%H & $2.9\pm0.5$ & $370\pm130$ & $580\pm200$ & $480 \pm 160 $ & $2.3 \pm 0.3$ & $0.78\pm0.14$ & $-1.20\pm0.84$ & 1.13 & 0.67& $1.83\pm0.61$ & $1.86\pm0.5$\\

\subsection{Ellipsoidal Cloud Virial Analysis}

Following Bertoldi \& McKee (1992, hereafter BM92), we consider an
ellipsoidal cloud with radius $R$ normal to the axis of symmetry and
size $2Z$ along the axis. The aspect ratio is defined as $y\equiv
Z/R$, while $R_{\rm max}$ and $R_{\rm min}$ are the semimajor and
semiminor axes of the ellipse obtained by projecting the cloud onto
the plane of the sky.

IRDCs F and H both have relatively thin, filamentary morphologies: we
set $R_{\rm max}/R_{\rm min}=3.40, 1.78$, respectively, i.e. the same
elongation as the rectangular regions we consider to define the
filaments (Figs. \ref{8panelF} \& \ref{8panelH}). Given these
morphologies, we expect the symmetry axes of the clouds to be close to
the plane of the sky. Thus for both we adopt a fiducial value of the
inclination angle between the cloud symmetry axis and the line of
sight of $\theta=60^\circ$. We assume $R=R_{\rm min}$ and $Z=R_{\rm
  max}/{\rm sin}\theta$, so $y=3.93,2.06$ for IRDCs F and H,
respectively. An uncertainty of $15^\circ$ in inclination would cause
$\sim 15\%$ uncertainties in $y$. It is also useful to introduce a
geometric mean observed radius, $R_{\rm obs}\equiv (R_{\rm max} R_{\rm
  min})^{1/2}$, which is also related to $R$ via, $R_{\rm obs}=R {\rm
  cos}^{1/2}\theta[1+(y {\rm tan} \theta)^2]^{1/4}$. BM92 also define
$R_m$ as the mean value of $R_{\rm obs}$ averaged over all viewing
angles, but for our individual clouds we will express quantities in
terms of $R$ and $R_{\rm obs}$. It should be noted that the treatment
of these IRDC filaments as simple ellipsoids is necessarily
approximate, and we consider the filamentary analysis of
section~\ref{S:fil} to be more accurate.

If the clump is in an environment that is evolving with a time scale
longer than the clump's dynamical time scale or signal crossing time
$t_s\equiv 2R/\sigma$, then it should obey the equilibrium virial
equation (McKee \& Zweibel 1992):
\beq
0 = 2({\cal T}-{\cal T}_0) + {\cal M} + W.
\label{eq:equilibriumvirial}
\eeq Here, ${\cal T}$ is the clump kinetic energy, ${\cal T}_0$ is the
kinetic energy resulting from the surface pressure on the clump,
${\cal M}$ is the magnetic energy associated with the cloud, and $W$
is the gravitational binding energy, which for an ellipsoidal cloud is
(BM92) \beq W = -\frac{3}{5} a_1 a_2 \frac{GM^2}{R}, \eeq where for a
power-law density distribution $\rho\propto r^{-k_\rho}$,
$a_1=(1-k_\rho/3)/(1-2k_\rho/5)$ and \beq a_2=\frac{{\rm arcsinh}
  (y^2-1)^{1/2}}{(y^2-1)^{1/2}} \eeq for prolate clouds. Note, our
definition of $W$ and $a_2$ differs slightly from BM92 since we do not
need to consider $R_m$. We adopt $k_\rho=1$ (based on a study of the
density profiles in IRDCs - Butler \& Tan, in prep.) so that
$a_1=10/9$. For our measured values of $y$, we have $a_2=0.53,0.71$
for IRDCs F and H, respectively. For the mass of the cloud we take the
average values of the estimates from MIR dust extinction and $\thco$
line emission. Using these values we estimate $W = - (11.0,1.08)\times
10^{46}\:{\rm erg}$ for IRDCs F and H, respectively (see Table 2). The
uncertainties in $W$ are relatively large given the measurement errors
of $\Sigma$ and $R$.

The clump kinetic energy is ${\cal T} = (3/2) M \sigma^2$, where
$\sigma$ is the average total 1D velocity dispersion, which we derive
from the $\thco$ line emission (counting only those parts of the
envelope-subtracted spectra with positive signal greater than or equal
to one standard deviation of the noise level) including corrections
for the molecular weight of $\thco$ and our adopted cloud
temperatures. We estimate that we measure $\sigma$ to a 10\%
accuracy. We find ${\cal T} = (16.4,2.06)\times 10^{46}\:{\rm erg}$
for IRDCs F and H, with uncertainties at about the 40\% level.

The surface term for the kinetic energy is ${\cal T}_0 = (3/2) P_0
V$. We estimate this term in two ways. First (method A), we can
measure the mass surface density of the surrounding molecular cloud
from the $\thco$ emission of the envelope regions. We find
$\Sigco({\rm env})= 0.0196, 0.0212\:{\rm g\:cm^{-2}}$ for IRDCs F and
H, respectively. We scale these values by $M/M_{\rm 13CO}$, i.e. 0.79
and 1.30.  If the envelope is self-gravitating, it has mean internal
pressure $P({\rm env})=1.85 G \Sigma^2({\rm env})$, adapting the
analysis of McKee \& Tan (2003) with parameters $f_g=1$ (i.e. fully
gas dominated), $\phi_{\rm geom}\equiv R_{\rm obs}^3/(R^2Z)=1.4$
(adopting an intermediate value for the two IRDCs accurate to about
20\%), $\phi_B=2.8$ (the fiducial value of McKee \& Tan, measuring the
ratio of the total pressure including magnetic fields to that assuming
they were absent) and $\alpha_{\rm vir}=1$. Thus setting $P_0 = P({\rm
  env})= (2.96,9.38)\times 10^{-11}$~cgs for IRDCs F and H and with
the cloud volume $V=4\pi R^2Z/3$, we find ${\cal
  T}_{0}(A)=(14,2.7)\times 10^{46}\:{\rm erg}$.

Second (method B), we estimate the density in the envelope region,
assuming it has a cylindrical, annular volume with outer radius $2R$
and inner radius $R$. For IRDCs F and H, we find densities of $\rho =
4 \Sigma({\rm env})/(3\pi R) = (1.43,3.20) \times 10^{-21}\:{\rm
  g\:cm^{-3}}$, equivalent to $n_{\rm H}({\rm env}) = (610,1600)\:{\rm
  cm^{-3}}$. We again scale these values by $M/M_{\rm 13CO}$,
i.e. 0.79 and 1.30 for IRDCs F and H, respectively.  We then equate
$P_0=\rho({\rm env}) \sigma^2({\rm env})$, where $\sigma({\rm env})$ is
the velocity dispersion of the envelope gas (we find $3.65,1.89\:{\rm
  km\:s^{-1}}$ for IRDCs F and H) and evaluate ${\cal T}_{0}(B) =
(3/2) V P_0 \rightarrow (73,4.3)\times 10^{46}\:{\rm erg}$, with
$V=(4/3)\pi R^2 Z$.

These results indicate that, for both IRDCs, the surface pressure term
of the virial equation is comparable to or much larger than the
internal kinetic term, although the uncertainties are large. Assuming
${\cal T}_0\geq{\cal T}$, then for virial equilbrium to be maintained
would require ${\cal M}\equiv (1/8\pi) \int_{V_a} (B^2-B_0^2)dV \geq
-W$, where $B_0$ is the magnetic field strength far from the cloud and
$V_a$ is a volume that extends beyond the cloud where the field lines
have been distorted by the cloud. Assuming $V_a$ is the volume of the
envelope regions and assuming negligible $B_0$, we find $B\geq
10,13\:{\rm \mu G}$ for IRDCs F and H. If a more realistic value of
$B_0=10\:{\rm \mu G}$ is adopted (Crutcher et al. 2010), then we find
$B\geq 14,16\:{\rm \mu G}$. Thus relatively modest magnetic field
enhancements could stabilize the clouds.

Bertoldi \& McKee (1992) define a dimensionless virial parameter,
$\alpha \equiv 5\sigma^2 R_m/(GM)$ to describe the dynamical state of
clouds. We adopt a slightly revised definition $\alpha \equiv
5\sigma^2 R/(GM) = a_1a_2 2 {\cal T}/|W|$, with $a_2$ defined as
above. For IRDCs F and H we find $\alpha=1.78, 3.17$. While these
values are quite close to unity, especially for IRDC F, which is
commonly taken to infer that self-gravity is important, this is
somewhat misleading since the value of $a_2$ is quite small for these
elongated clouds and the surface pressure terms seem to be quite
important. 

It is possible that these IRDCs have not yet reached virial
equilibrium if their surroundings are evolving on timescales $\lesssim
t_s\sim$ a few Myr. The mean velocity of the $\thco$ emitting gas in
the north and south envelope regions is 55.72, 56.10~$\kms$,
respectively, for IRDC F, and 44.96, 44.28~$\kms$, respectively, for
IRDC H. The north/south velocity dispersions are 3.81/3.38~$\kms$ for
F and 1.69/2.07~$\kms$ for H. The north/south $\Sigco$'s are
0.040/0.031~$\rm g\:cm^{-2}$ for F and 0.050/0.051~$\rm g\:cm^{-2}$
for H. The ratios of north/south pressures (estimated by method B,
$\propto \Sigma\sigma^2$) are thus 1.64 and 0.65 for IRDCs F and
H. The pressures appear to be fairly similar on the different sides of
the filaments, although the uncertainties are such that variation at
the level of about a factor of two could be present.

%emulate
\newpage

\subsection{Filamentary Cloud Virial Analysis}\label{S:fil}

Fiege \& Pudritz (2000, hereafter FP00) present a virial analysis of
filamentary clouds. They derived the following equation satisfied by
pressure-confined, nonrotating, self-gravitating,
filamentary (i.e. lengths $\gg$ widths) clouds threaded by helical
magnetic fields that are in virial equilibrium:
\begin{equation}
\frac{P_0}{P}=1-\frac{m}{m_{\rm vir}} \left(1-\frac{{\cal M}_l}{|W_l|}\right).
\label{cylvireqn}
\end{equation}
%(Eq.~\ref{cylvireqn})
Here $P_0$ is the external pressure at the surface of the filament,
$P = \rho \sigma^2$ is the average total pressure in the filament, $m$
is the mass per unit length, $m_{\rm vir}\equiv 2
\sigma^2 / G$ is the virial mass per unit length, ${\cal
M}_l$ is the magnetic energy per unit length, and $W_l=-m^2G$ is the
gravitational energy per unit length.

We divide IRDCs F, H into 7, 4 orthogonal strips (see
Figs.~\ref{8panelF} \& \ref{8panelH}) with angular widths 1.70\arcmin,
0.955\arcmin\ along the filaments, respectively. Assuming a fiducial
value of the inclination angle between the cloud symmetry axis and the
line of sight of $\theta=60^\circ$, these correspond to physical
lengths along the filaments of 2.11, 0.930~pc, respectively.

In Table~\ref{tab:filvir}, for each strip in IRDCs F and H, we list the
values of $\Sigco$, $\Sigsmf$, $M$ (calculated from the mean of these
values of $\Sigma$), $m$, $\rho$, $\bar{v}$, $\sigma$, $m_{\rm vir}$,
$P$, $\Sigco({\rm env})$, $\rho({\rm env})$ (calculated after scaling
$\Sigco({\rm env})$ by $0.5(\Sigco + \Sigsmf)/\Sigco$), $\bar{v}({\rm
  env})$, $\sigma({\rm env})$ and $P({\rm env})\equiv \rho({\rm
  env})\sigma^2({\rm env})$. We equate $P_0=P({\rm env})$.

\begin{figure*}
\begin{center}
\includegraphics[width=5in,angle=0]{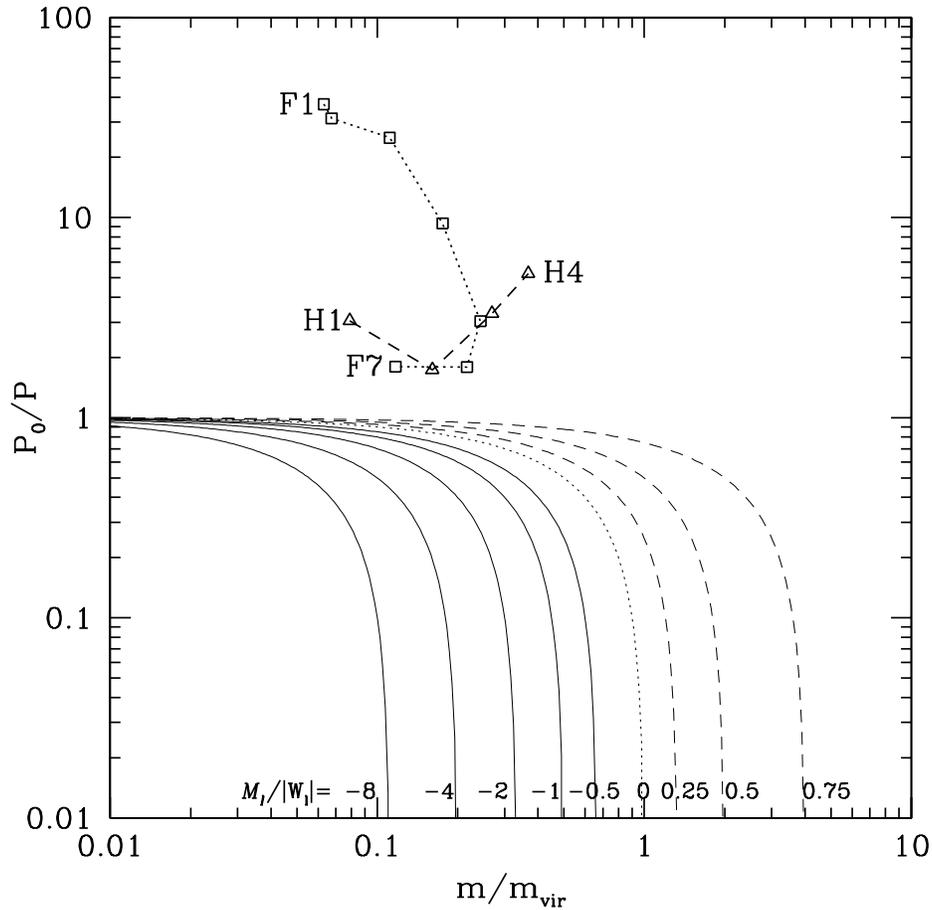}
\end{center}
\caption{$P_0/P$ versus $m/m_{\rm vir}$ for strips in IRDCs F (open squares joined by dotted line for F1 to F7) and H (open triangles joined by dashed line for H1 to H4). The smooth curves show the conditions satisfied by equation~(\ref{cylvireqn}) for ${\cal M}_l/|W_l|<0$ (solid lines), ${\cal M}_l/|W_l|=0$ (dotted line), and ${\cal M}_l/|W_l|>0$ (dashed lines). The large observed values of $P_0/P$ may indicate that most of the regions in IRDCs F and H have not yet established virial equilibrium. Those strips with the lowest values of $P_0/P$, i.e. F6, F7, and H2, appear to be undergoing more active star formation (see text and Figs.~\ref{8panelF} \& \ref{8panelH}).
}
\label{fig:fiege}
\end{figure*}

Following FP00, in Fig.~\ref{fig:fiege} we plot $P_0/P$ versus
$m/m_{\rm vir}$. The range of models considered by FP00 allows for
positive values of ${\cal M}_l/|W_l|$ (i.e. poloidally-dominated
B-fields that provide net support to the filament against
gravitational collapse) and negative values (i.e. toroidally-dominated
B-fields that provide net confinement of the filament). In all cases,
$P_0/P\leq 1$. In contrast, we find all of the filament
regions have $P_0/P>1$, i.e. the pressures in the envelope regions
appear to be greater than in the filament. This echoes the results
from the ellipsoidal virial analysis, which found large surface
pressure terms. Assuming our measurements of pressures are reliable,
e.g. are not being adversely affected by systematic effects due to, for
example, our assumed filament and envelope geometry, then these
results imply that the filaments have not yet reached virial equilibrium.

Very large values of $P_0/P$ are inferred for strips F1, F2 and
F3. These are consistent with the filament and envelope spectra shown
in Fig.~\ref{envspectra}, which reveal a relatively weak filament and
relatively strong and varying envelope velocity profiles.

Strips F6, F7 and H2 have the smallest values of $P_0/P\lesssim 2$.
Examining the IRAC $\rm 8\mu m$ images (Figs.~\ref{8panelF} \&
\ref{8panelH}), there is some indication that these are the sites of
relatively active star formation (especially F7 and H2). Star
formation requires gravitationally unstable conditions in the
filament, i.e. regions where self-gravity starts to dominate over
external pressure. Our results indicate that this also requires the
local region of the filament to reach approximate virial equilibrium,
although surface pressure terms still remain dynamically important,
i.e. $m/m_{\rm vir}$ is significantly less than unity. Given our
measurement uncertainties and the fact that the observed regions do
not have $P_0/P<1$, we are not able to determine whether the field
geometries are more dominated by poloidal or toroidal components. We
note that FP00's conclusion that observed filaments are dominated by
toroidal fields depends on their assumption that the filaments are in
virial equilibrium and on their choice of $P_0$, which was not
directly measured for most of the sources they considered.

We caution that if independent molecular clouds are present along
these lines of sight and with similar velocities to the filaments,
then this may cause us to overestimate the velocity dispersion and
pressure in the envelope regions around the filaments. From the
spectra shown in Fig.~\ref{envspectra} we do not expect this is
occurring in IRDC H, since the envelope spectra share a very similar
velocity range as the filament. The situation in IRDC F is less clear
cut, since there appears to be a broader, offset component that
dominates more in the envelope region. 
%For IRDC F, we have
%experimented with varying the velocity interval used to define the
%filament and envelope, in particular considering a narrower range more
%centered on the envelope-subtracted filament at $\sim 55-60\:{\rm
%  km\:s^{-1}}$.
%We considered a narrower velocity range of 58-62~$\rm km\:s^{-1}$ for
%strips F1 and F2, and 57-62~$\rm km\:s^{-1}$ for F3.

\begin{deluxetable}{c|cccccccc|ccccc}
%emulate
%\rotate
%\begin{deluxetable*}{lccccc}
\tabletypesize{\footnotesize}
%\tablecolumns{5}
\tablewidth{0pt}
\tablecaption{IRDC Filamentary Virial Analysis}
\tablehead{\colhead{Cloud property} &
           \colhead{F1} &
           \colhead{F2} &
           \colhead{F3} &
           \colhead{F4} &
           \colhead{F5} &
           \colhead{F6} &
           \colhead{F7} &
           \colhead{$\rm F_{\rm tot}$} & 
           \colhead{H1} &
           \colhead{H2} &
           \colhead{H3} &
           \colhead{H4} &
           \colhead{$\rm H_{\rm tot}$} 
}
\startdata
$\Sigma_{\rm 13CO}$ ($\rm 10^{-2}\:g\:cm^{-2}$) & 0.218 & 0.338 & 0.811 & 1.31 & 2.29 & 2.87 & 2.58 & 1.47 & 0.864 & 1.67 & 1.56 & 1.21 & 1.33\\
$\Sigma_{\rm SMF}$ ($\rm 10^{-2}\:g\:cm^{-2}$) & 0.476 & 0.311 & -0.0432 & 0.862 & 1.37 & 1.62 & 1.03 & 0.823 & 1.38 & 1.94 & 1.97 & 3.28 & 2.09\\
$M$ ($M_\odot$) & 114 & 107 & 126 & 357 & 601 & 737 & 593 & 2640 & 78.5 & 126 & 124 & 157 & 478\\
$m$ ($M_\odot {\rm pc^{-1}}$) & 54.0 & 50.7 & 59.7 & 169 & 285 & 349 & 281 & 178 & 84.4 & 135 & 133 & 169 & 128\\
$\rho$ ($10^{-22}{\rm g\:cm^{-3}}$) & 3.29 & 3.09 & 3.64 & 10.3 & 17.4 & 21.3 & 17.1 & 10.9 & 22.0 & 35.1 & 34.6 & 44.0 & 33.3\\
$\bar{v}$ (${\rm km\:s^{-1}}$) & 59.19 & 59.34 & 58.81 & 58.26 & 58.30 & 57.77 & 57.77 & 58.39 & 45.09 & 45.20 & 45.13 & 44.78 & 45.07\\
$\sigma$ (${\rm km\:s^{-1}}$) & 1.36 & 1.27 & 1.07 & 1.44 & 1.59 & 1.86 & 2.27 & 1.46 & 1.52 & 1.34 & 1.03 & 0.995 & 1.20\\
$m_{\rm vir}$ ($M_\odot {\rm pc^{-1}}$) & 856 & 751 & 536 & 962 & 1170 & 1610 & 2400 & 986 & 1070 & 840 & 494 & 460 & 669\\
$P$ ($10^{-12}{\rm cgs}$) & 6.05 & 4.99 & 4.19 & 21.3 & 43.8 & 73.8 & 88.3 & 23.1 & 50.7 & 63.4 & 36.8 & 43.6 & 47.9\\
$\Sigma_{\rm 13CO}({\rm env})$ ($\rm 10^{-2}\:g\:cm^{-2}$) & 1.61 & 1.83 & 1.94 & 2.14 & 1.81 & 1.95 & 2.46 & 1.96 & 2.31 & 1.86 & 2.00 & 2.35 & 2.12 \\
$\rho({\rm env})$ ($\rm 10^{-22}\:g\:cm^{-3}$) & 18.7 & 12.9 & 6.72 & 13.0 & 10.6 & 11.2 & 12.6 & 11.2 & 45.3 & 30.4 & 34.2 & 65.9 & 41.2 \\
$\bar{v}({\rm env})$ (${\rm km\:s^{-1}}$) & 56.03 & 55.84 & 55.78 & 56.59 & 56.04 & 55.65 & 55.72 & 55.95 & 44.97 & 44.78 & 44.53 & 44.40 & 44.67\\
$\sigma$({\rm env}) (${\rm km\:s^{-1}}$) & 3.45 & 3.48 & 3.95 & 3.91 & 3.55 & 3.43 & 3.56 & 3.65 & 1.85 & 1.91 & 1.89 & 1.86 & 1.89\\
%$\sigma$({\rm env}) (${\rm km\:s^{-1}}$) & 0.917 & 1.105 & 1.46 & 1.90 & 1.76 & 1.79 & 1.64 & 1.78 & 1.68 & 1.70 & 1.70 & 1.64 & 1.70\\
$P({\rm env})$ ($10^{-12}{\rm cgs}$) & 223 & 156 & 105 & 199 & 133 & 132 & 159 & 149 & 155 & 110 & 122 & 229 & 148\\

%$\bar{\Sigma}_{\rm SMF}({\rm fil})$ ($\rm g\:cm^{-2}$)& $0.0082\pm0.0016$ & $0.0209\pm0.004$\\

%$\bar{\Sigma}_{\rm 13CO}({\rm env})$ ($\rm g\:cm^{-2}$) & $0.0196\pm0.004$ & $0.0212\pm0.004$\\
%$M_{\rm 13CO}$ ($M_\odot$) & $3300\pm1100$ & $370\pm150$\\
%$M_{\rm SMF}$ ($M_\odot$)& $1900\pm640$ & $580\pm230$\\
%$M$ ($M_\odot$)& $2600\pm 870$ & $480 \pm 190$\\
%$a_1$ & 10/9 & 10/9\\
%$a_2$ & 0.538 & 0.750\\
%$W$ ($10^{46}$erg)& $-11.0\pm7.1$& $-1.08\pm0.78$\\
%$\sigma$ (km/s) & 1.27 & 1.13\\
%$t_s=2R/\sigma$ (Myr) & $2.9\pm0.37$ & $1.6\pm0.27$ \\
%${\cal T}$ ($10^{46}$erg) & $12.5\pm4.2$ & $1.83\pm0.73$\\
%${\cal T}_0(A)$ ($10^{46}$erg) & $14\pm8$ & $2.7\pm1.8$\\
%${\cal T}_0(B)$ ($10^{46}$erg) & $55\pm17$ & $9.9\pm3.5$\\
%$\alpha$ & $1.36\pm 0.33$ & $2.81\pm 0.74$\\
\enddata
\label{tab:filvir}
\end{deluxetable}

\section{Conclusions}

We have compared measurements of mass surface density, $\Sigma$, in
two IRDC filaments based on $\thco$ observations, $\Sigco$, with those
derived from MIR extinction mapping, $\Sigsmf$, finding agreement at
the factor of $\sim 2$ level. A systematic decrease in
$\Sigco/\Sigsmf$ with increasing $\Sigma$ may be due to a systematic
decrease in temperature, increase in the contribution of
under-resolved high optical depth regions, increase in dust opacity,
or decrease in $\thco$ abundance due to depletion in regions of higher
column density. Future studies that spatially resolve the temperature
structure and MIR dust absorption properties can help to distinguish
these possibilities.

We have then used the kinematic information derived from $\thco$ to
study the dynamical state of the IRDCs. In particular we have
evaluated the terms of the steady-state virial equation, including
surface terms, under the assumption of ellipsoidal and filamentary
geometries. In both cases we find evidence that the surface pressure
terms are important and possibly dominant, which may indicate that the
filaments, at least globally, have not yet reached virial
equilibrium. 

These results would be consistent with models of compression of dense
gas in colliding molecular flows, e.g. GMC collisions. Tan (2000)
proposed that this mechanism may trigger the majority of star
formation in shearing disk galaxies. The expected collision velocities
are $\sim 10\:{\rm km\:s^{-1}}$. It is less clear whether colliding
atomic flows, (e.g. Heitsch et al. 2008), which form the molecular gas
after shock compression of atomic gas, would also produce such
kinematic signatures: recall that we are inferring large surface
pressures based on $\thco$ emission from the envelopes around the IRDC
filaments.

Recent observations of extended, parsec-scale SiO emission, likely
produced in shocks with velocities $\gtrsim 12\:{\rm km\:s^{-1}}$ in
IRDC H by Jim\'enez-Serra et al. (2010) may also support models of
filament formation from converging flows. However, we caution that the
observed extended SiO emission is very weak and may also be produced
by multiple protostellar outflow sources forming in the IRDC (see
Jim\'enez-Serra et al. 2010 for further discussion).

Our resolved filamentary virial analysis also indicates that the
regions closest to virial equilibrium (strips F6, F7 and H2) are those
which have initiated the most active star formation. This would be
expected if models of slow, equilibrium star formation (Tan et
al. 2006; Krumholz \& Tan 2007) apply locally in these regions. In
this case, these dense regions that have become gravitationally
unstable, perhaps due to the action of external pressure and/or
converging flows, then persist for more than one local dynamical time
and so are able to reach approximate pressure and virial equilibrium
with their surroundings. In this scenario, they are stabilized by the
ram pressure generated by protostellar outflow feedback from the
forming stars (Nakamura \& Li 2007).

\acknowledgements We thank Michael Butler for providing the MIR
extinction maps used in this analysis. We also thank Peter Barnes,
Crystal Brogan, Michael Butler, Paola Caselli, James Jackson, Izaskun
Jim\'enez-Serra, Thushara Pillai and Robert Simon for helpful
discussions. The comments of an anonymous referee led to improvements
in the paper. AKH acknowleges support from a SEAGEP Dissertation
Fellowship. JCT acknowledges support from NSF CAREER grant AST-0645412
and NASA Astrophysics Theory and Fundamental Physics grant ATP09-0094.
 
%\begin{references}

%\reference{} Ao, Y., Yang, J., Sunada, K. 2004, \apj, 128, 1716 
%\reference{} Benjamin, R. A., Churchwell, E., Babler, B. L., Bania, T. M., Clemens, D. P. et al. 2003, \pasp, 115, 953

%\reference{} Battersby, C., Bally, J., Jackson, J. M. et al. 2010, \apj, 721, 222

\end{document}